\documentclass[fleqn,usenatbib]{mnras}

\usepackage{newtxtext,newtxmath}
\usepackage[T1]{fontenc}
\usepackage{ae,aecompl}
\usepackage{longtable}

\usepackage[english]{babel}
\usepackage{graphicx}

\newcommand{\HeI}{\mbox{He\hspace{0.25em}{\sc i}}}
\newcommand{\HeII}{\mbox{He\hspace{0.25em}{\sc ii}}}

\newcommand{\OI}{\mbox{O\hspace{0.25em}{\sc i}}}

\newcommand{\CaII}{\mbox{Ca\hspace{0.25em}{\sc ii}}}

\newcommand{\Nifs}{$^{56}$Ni}

\title[]{The Fast, Luminous Ultraviolet Transient AT2018cow: Extreme Supernova, or Disruption of a Star by an Intermediate-Mass Black Hole?}

\newcommand{\ljmu}{1}
\newcommand{\mpa}{2}
\newcommand{\caltech}{3}
\newcommand{\gsfc}{4}
\newcommand{\umd}{5}
\newcommand{\ucsc}{6}
\newcommand{\dark}{7}
\newcommand{\iia}{8}
\newcommand{\iis}{9}
\newcommand{\ipmu}{10}
\newcommand{\konan}{11}
\newcommand{\tohoku}{12}
\newcommand{\unam}{13}
\newcommand{\fsu}{14}
\newcommand{\iitb}{15}
\newcommand{\ucb}{16}
\newcommand{\asu}{17}
\newcommand{\coo}{18}
\newcommand{\suphys}{19}
\newcommand{\sdsu}{20}
\newcommand{\tit}{21}
\newcommand{\kyoto}{22}
\newcommand{\cddd}{23}
\newcommand{\ncugrad}{24}
\newcommand{\soka}{25}
\newcommand{\iasf}{26}
\newcommand{\nbi}{27}
\newcommand{\ensenada}{28}
\newcommand{\suastro}{29}
\newcommand{\yalenus}{30}

\author[Perley et al.]{Daniel A. Perley,$^{\ljmu}$
Paolo A. Mazzali,$^{\ljmu,\mpa}$
Lin Yan,$^{\caltech}$
S. Bradley Cenko,$^{\gsfc,\umd}$
Suvi Gezari,$^{\umd}$
\newauthor
Kirsty Taggart,$^{\ljmu}$
Nadia Blagorodnova,$^{\caltech}$
Christoffer Fremling,$^{\caltech}$
Brenna Mockler,$^{\ucsc,\dark}$
\newauthor
Avinash Singh,$^{\iia,\iis}$
Nozomu Tominaga,$^{\ipmu,\konan}$
Masaomi Tanaka,$^{\tohoku}$
Alan M. Watson,$^{\unam}$
\newauthor
Tom{\'a}s Ahumada,$^{\umd}$
G. C. Anupama,$^{\iia}$
Chris Ashall,$^{\fsu}$
Rosa L. Becerra,$^{\unam}$
\newauthor
David Bersier,$^{\ljmu}$
Varun Bhalerao,$^{\iitb}$
Joshua S. Bloom,$^{\ucb}$
Nathaniel R. Butler,$^{\asu}$
\newauthor
Chris Copperwheat,$^{\ljmu}$
Michael W. Coughlin,$^{\caltech}$
Kishalay De,$^{\caltech}$
Andrew J. Drake,$^{\caltech}$
\newauthor
Dmitry A. Duev,$^{\caltech}$
Sara Frederick,$^{\umd}$
J. Jes\'us Gonz\'alez,$^{\unam}$
Ariel Goobar,$^{\suphys}$
\newauthor
Marianne Heida,$^{\caltech}$
Anna Y. Q. Ho,$^{\caltech}$
John Horst,$^{\sdsu}$
Tiara Hung,$^{\umd}$
Ryosuke Itoh,$^{\tit}$
\newauthor
Jacob E. Jencson,$^{\caltech}$
Mansi M. Kasliwal,$^{\caltech}$
Nobuyuki Kawai,$^{\tit}$
Tanazza Khanam,$^{\iitb}$
\newauthor
Shrinivas R. Kulkarni,$^{\caltech,\coo}$
Brajesh Kumar,$^{\iia}$
Harsh Kumar,$^{\iitb}$
Alexander S. Kutyrev,$^{\gsfc,\umd}$
\newauthor
William H. Lee,$^{\unam}$
Keiichi Maeda,$^{\kyoto}$
Ashish Mahabal,$^{\caltech,\cddd}$
Katsuhiro L. Murata,$^{\tit}$
\newauthor
James D. Neill,$^{\caltech}$
Chow-Choong Ngeow,$^{\ncugrad}$
Bryan Penprase,$^{\soka}$
Elena Pian,$^{\iasf}$
\newauthor
Robert Quimby,$^{\sdsu,\ipmu}$
Enrico Ramirez-Ruiz,$^{\ucsc,\nbi}$
Michael G. Richer,$^{\unam}$
\newauthor
Carlos G. Rom\'an-Z\'u\~niga,$^{\ensenada}$
D. K. Sahu,$^{\iia}$
Shubham Srivastav,$^{\iitb}$
Quentin Socia,$^{\sdsu}$
\newauthor
Jesper Sollerman,$^{\suastro}$
Yutaro Tachibana,$^{\tit}$
Francesco Taddia,$^{\suastro}$
Samaporn Tinyanont,$^{\caltech}$
\newauthor
Eleonora Troja,$^{\gsfc,\umd}$
Charlotte Ward,$^{\umd}$
Jerrick Wee,$^{\yalenus}$
Po-Chieh Yu$^{\ncugrad}$
\\
$^{\ljmu}$ Astrophysics Research Institute, Liverpool John Moores University, 146 Brownlow Hill, Liverpool L3 5RF, UK\\
$^{\mpa}$ Max-Planck Institut f\"ur Astrophysik, Karl-Schwarzschild-Str. 1, 84741 Garching, Germany \\
$^{\caltech}$ Division of Physics, Math, and Astronomy, California Institute of Technology, Pasadena, CA 91125, USA\\
$^{\gsfc}$ Astrophysics Science Division, NASA Goddard Space Flight Center, 8800 Greenbelt Road, Greenbelt, MD 20771, USA \\
$^{\ucsc}$ Department of Astronomy and Astrophysics, University of California, Santa Cruz, CA 95064, USA \\
$^{\dark}$ Dark Cosmology Centre, Niels Bohr Institute, University of Copenhagen, Blegdamsvej 17, 2100 Copenhagen, Denmark \\
$^{\umd}$ Department of Astronomy, University of Maryland, College Park, MD 20742, USA \\
$^{\iia}$ Indian Institute of Astrophysics, II Block Koramangala, Bengaluru 560034, India \\
$^{\iis}$ Joint Astronomy Programme, Department of Physics, Indian Institute of Science, Bengaluru 560012, India\\
$^{\ipmu}$ Kavli IPMU (WPI),
  The University of Tokyo Institutes for Advanced Study,
  The University of Tokyo,
  Kashiwa, Chiba 277-8583, Japan\\
$^{\konan}$ Department of Physics, Faculty of Science and Engineering, Konan University, 8-9-1 Okamoto, Kobe, Hyogo 658-8501, Japan\\
$^{\tohoku}$ Astronomical Institute, Tohoku University, Aoba, Sendai 980-8578, Japan \\
$^{\unam}$ Instituto de Astronom\'ia, Universidad Nacional Aut\'onoma de M\'exico, Apartado Postal 70-264, 04510 M\'exico, CDMX, M\'exico \\
$^{\fsu}$ Department of Physics, Florida State Universiy, Tallahassee, FL 32306, USA\\
$^{\iitb}$ Physics Department, Indian Institute of Technology Bombay, Powai, Mumbai 400076, India\\
$^{\ucb}$ Department of Astronomy, University of California, Berkeley, 94720, USA \\
$^{\asu}$ School of Earth and Space Exploration, Arizona State University, Tempe, AZ 85287, USA \\
$^{\coo}$ Caltech Optical Observatories, California Institute of Technology, Pasadena, CA 91125, USA\\
$^{\suphys}$ The Oskar Klein Centre, Department of Physics, AlbaNova, Stockholm University, SE-106 91 Stockholm, Sweden\\
$^{\sdsu}$ Department of Astronomy/Mount Laguna Observatory, San Diego State University, 5500 Campanile Drive, San Diego, CA 92812-1221, USA \\
$^{\tit}$ Department of Physics, School of Science, Tokyo Institute of Technology, 2-12-1 Ohokayama, Meguro, Tokyo 152-8551, Japan \\
$^{\kyoto}$ Department of Astronomy, Kyoto University, Kitashirakawa-Oiwake-cho, Sakyo-ku, Kyoto 606-8502, Japan \\
$^{\cddd}$ Center for Data Driven Discovery, California Institute of Technology, Pasadena, CA 91125, USA\\
$^{\ncugrad}$ Graduate Institute of Astronomy, National Central University, 32001, Taiwan \\
$^{\soka}$ Soka University of America, 1 University Drive, Aliso Viejo, CA 92656, USA \\
$^{\iasf}$ INAF OAS, Via Piero Gobetti, 101, I-40129 Bologna, Italy\\
$^{\nbi}$ Niels Bohr Institute, University of Copenhagen, Blegdamsvej 17, 2100 Copenhagen, Denmark \\
$^{\ensenada}$ Instituto de Astronom\'ia UNAM, Unidad Acad\'emica en Ensenada, Ensenada BC, 22860, M\'exico \\
$^{\suastro}$ The Oskar Klein Centre, Department of Astronomy, AlbaNova, Stockholm University, SE-106 91 Stockholm, Sweden\\
$^{\yalenus}$  Yale-NUS College, 16 College Avenue West, Singapore 138527, Singapore
}

\date{}

\pubyear{2018}

\begin{document}
\label{firstpage}
%\pagerange{\pageref{firstpage}--\pageref{lastpage}}

\maketitle

\begin{abstract}
Wide-field optical surveys have begun to uncover large samples of fast ($t_{\rm rise} \lesssim 5$d), luminous ($M_{\rm peak} < -18$), blue transients.  While commonly attributed to the breakout of a supernova shock into a dense wind, the great distances to the transients of this class found so far have hampered detailed investigation of their properties.
We present photometry and spectroscopy from a comprehensive worldwide campaign to observe AT\,2018cow (ATLAS\,18qqn), the first fast-luminous optical transient to be found in real time at low redshift.  Our first spectra ($<2$ days after discovery) are entirely featureless.  A very broad absorption feature suggestive of near-relativistic velocities develops between $3-8$ days, then disappears.  Broad emission features of H and He develop after $>10$ days.  The spectrum remains extremely hot throughout its evolution, and the photospheric radius contracts with time (receding below $R<10^{14}$ cm after 1 month).  This behaviour does not match that of any known supernova, although a relativistic jet within a fallback supernova could explain some of the observed features.   Alternatively, the transient could originate from the disruption of a star by an intermediate-mass black hole, although this would require long-lasting emission of highly super-Eddington thermal radiation.  In either case, AT\,2018cow suggests that the population of fast luminous transients represents a new class of astrophysical event. Intensive follow-up of this event in its late phases, and of any future events found at comparable distance, will be essential to better constrain their origins.
\end{abstract}

\section{Introduction}

The development of sensitive, wide-area digital optical sky surveys has led to the discovery of populations of rare, luminous extragalactic transients that evolve on timescales of just a few days---much faster than typical supernovae, whose light curves are governed by the decay of \Nifs\ within a massive envelope and typically take weeks to months to fade.  Many of these have been reasonably well-explained by known phenomena: shock-breakout flashes from supernovae \citep[e.g.,][]{Ofek+2010,Shivvers+2016,Arcavi+2017}, early emission from relativistic supernovae \citep{Whitesides+2017}, or the shockwave afterglows from gamma-ray bursts \citep{Cenko+2013,Cenko+2015,Stalder+2017,Bhalerao+2017}.  

Other objects are more mysterious, however, and still lack a convincing explanation or firm spectroscopic identification.  In particular, populations of optical transients with luminosities comparable to or exceeding those of the most luminous core-collapse supernovae, but rise times of only a few days, have been reported by a variety of different surveys \citep{Arcavi+2016,Drout+2014,Tanaka+2016,Pursiainen+2018,Rest+2018}.  Nearly all of these events (dubbed fast-evolving luminous transients by \citealt{Rest+2018}) were found at great distances ($z>0.1$) where they are difficult to study.  Furthermore most were not recognized as unusual events in real time, preventing the acquisition of essential follow-up observations.  The few spectra that are available tend to show only featureless blue continuua.  Because of their origins in star-forming galaxies these transients are widely interpreted as supernovae, but strong constraints are lacking.

Fortunately, our ability to find and identify fast transients continues to improve, and several surveys are now monitoring almost the entire sky at cadences of a few days or less.  The Asteroid Terrestrial-impact Last Alert System (ATLAS; \citealt{Tonry+2018}) observes most of the visible Northern sky down to 19 mag every $\sim2$ nights.   The Zwicky Transient Facility (ZTF; \citealt{ATEL11266}) observes a similar area to 20.5 mag every 3 nights, and a significant fraction of it at much higher cadence.  ASAS-SN \citep{Shappee+2014} monitors both hemispheres nightly to $\sim$17 mag.  With these capabilities, it is now possible to find and identify transients in (almost) real time over most of the night sky.

In this paper, we present a detailed observational study of the first fast high-luminosity transient to be identified in the nearby universe in real time: AT\,2018cow, discovered by the ATLAS survey and independently detected by ZTF and ASAS-SN.   We present our extensive, worldwide observational campaign in \S \ref{sec:observations}, focusing on observations at ultraviolet, optical, and near-infrared wavelengths (the multiwavelength view of this transient is presented by \citealt{Ho+2018}).   We summarize the key properties of this event in \S \ref{sec:properties}, and illustrate the ways in which AT\,2018cow is distinct from any well-established class of transient in \S \ref{sec:comparisons}.  In \S \ref{sec:interpretation} we consider two possible explanations for its origin: a jet-driven supernova erupting into a dense envelope of circumstellar matter, or alternatively the tidal disruption of a star around an intermediate-mass black hole located in a small galaxy's spiral arm.  Both models have significant difficulties explaining the full suite of observations, and our observations suggest that the origins of fast luminous transients may be significantly more exotic and complex than previously assumed.  We summarize our results and examine future directions in fast-transient research in \S \ref{sec:conclusions}.

\section{Observations}
\label{sec:observations}

\subsection{Discovery and Pre-Imaging Constraints}

AT2018cow\footnote{The name of this transient was assigned automatically by the Transient Name Server (https://wis-tns.weizmann.ac.il/).  It was later redesignated SN2018cow following the emergence of broad features in the spectrum, although we argue here that a SN association is not definite and retain the AT designation.  The transient is also known as ATLAS18qqn and as ZTF18abcfcoo.} was discovered and promptly announced via the Astronomers Telegram \citep{ATEL11727} by ATLAS; the discovery and early data are described in detail by \citealt{Prentice+2018}.  The first detection of the transient was an image taken at 2018-06-16 10:35:02 UT (MJD 58285.441), appearing as a strikingly bright (14.7 $\pm$ 0.1 mag in the ATLAS $o$-band) optical source coincident with the galaxy CGCG 137-068 ($z$ = 0.0141, $d$ = 60 Mpc\footnote{We assume $h$ = 0.7, $\Omega_M$ = 0.3, $\Omega_\Lambda$ = 0.7.}; \citealt{Abolfathi+2018}).  The preceding ATLAS observation of the field, four days earlier (MJD 58281.48), registered no detection of any transient object at the same location to a magnitude limit of $o > 20.2$ mag, implying brightening by almost 5 mag within this period.  Independent imaging by the Palomar 48-inch telescope (P48) as part of the ZTF public Northern Sky Survey later moved the time of last non-detection one day closer, to only three days before the first ATLAS detection ($i > 19.5$ at MJD 58282.172; \citealt{ATEL11738}).  The ASAS-SN non-detection reported by \citealt{Prentice+2018} ($g > 18.9$ at MJD 58284.13) provides an even tighter constraint: a rise of $>4.2$ magnitudes over $<$1.3 days. 

A fast rise to a very high optical luminosity ($M < -19$ mag) is unusual for supernovae but similar to cosmological fast-transients of the types discussed in the introductory paragraph.  Motivated by these unusual characteristics, we initiated a campaign of observations via the GROWTH (Global Relay of Observatories Watching Transients Happen) network, a world-wide collaboration of predominantly small telescopes co-operating in the study of energetic time-domain phenomena.  We also observed it under other telescopic programs.  Our observing campaign is described in detail below.

% Imaging figure
\begin{figure*}
\includegraphics[width=6.7in]{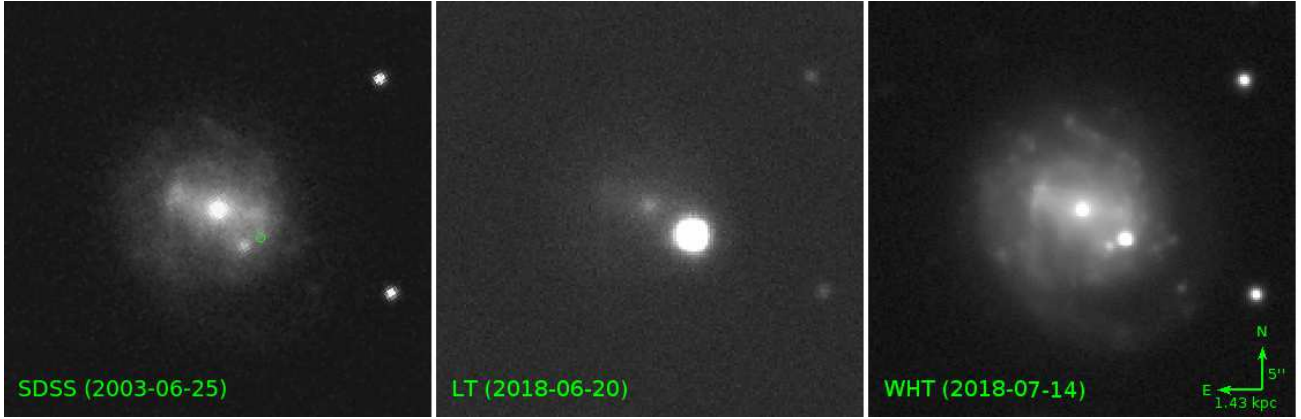}
    \caption{Pre-explosion imaging of AT\,2018cow from the Sloan Digital Sky Survey compared to imaging with the Liverpool Telescope taken shortly after peak and deep later-time imaging from the William Herschel Telescope.  The transient is significantly brighter than its host galaxy at peak.  The galaxy itself shows a barred morphology and weak spiral features, one of which underlies the transient.  A point-source located at the galaxy nucleus is likely to be a weak AGN, while a fainter compact source slightly southeast of the transient is likely an H II region.  No point source lies under the transient itself (position designated by a green circle in left panel), and there are no obvious merger indicators.}
    \label{fig:image}
\end{figure*}

\subsection{Ground-Based Imaging Observations}

Nightly imaging observations were acquired with the Infrared-Optical imager on the robotic Liverpool Telescope (LT; \citealt{Steele+2004}) in both optical (IO:O) and near-infrared  (IO:I) bands.  We typically observed with the full suite of available filters ($uBgVrizH$) although on some nights a more limited set was obtained.
We also obtained frequent imaging from a variety of other facilities.  These include the CCD imager on the Mount Laguna Observatory (MLO; \citealt{Smith+1969}) 1m telescope, the EMCCD demonstrator camera on the Kitt Peak 84-inch telescope (KP84), ANDICAM on the 1.5m telescope at the Cerro Tololo Interamerican Observatory, the Himalayan Faint Object Spectrograph Camera (HFOSC) on the 2-m Himalayan Chandra Telescope (HCT), the COATLI 50-cm Telescope \citep{Watson+2016} at the Observatorio Astron\'omico Nacional in Sierra San Pedro M\'artir, and the Reionization and Transients Infrared instrument (RATIR; \citealt{Butler+2012,Watson+2012}) on the 1.5-meter Harold L. Johnson telescope (also at San Pedro M\'artir).  Observations were taken less regularly with the 0.4m (SLT) and 1.0m (LOT) telescopes at Lulin Observatory in Taiwan, the MITSuME 50 cm telescope of Akeno Observatory in Japan, and with the Wide-Field Infrared Camera (WIRC) at the Palomar 200-inch Hale Telescope.  Finally, a single epoch of deep $r$-band imaging was acquired using the Auxiliary Port Camera (ACAM) on the William Herschel Telescope.

Images were reduced using standard methods.  A dithered sequence of NIR frames was not available for the ANDICAM NIR images and simple pair subtraction was used to remove the sky.

Host galaxy contribution to the transient flux is not insignificant (especially at late times; Figure \ref{fig:image}).  We used a custom image-subtraction tool written in IDL to remove the host galaxy flux from all ground-based optical images consistently by convolving both the transient image and a template image to a common PSF, then subtracting.  Imaging from the Sloan Digital Sky Survey (SDSS; \citealt{Abolfathi+2018}) was used to subtract the $ugriz$ measurements.  For non-SDSS optical filters ($UBVRI$) we averaged two adjacent filters: e.g., to simulate a $B$-band image we took a weighted average of the aligned $u$ and $g$ images.   The relative weights for each synthetic filter were estimated based on the relative magnitude weights from the Lupton transformation equations.\footnote{http://www.sdss3.org/dr8/algorithms/sdssUBVRITransform.php}

Host subtraction for the NIR images is more challenging: the only available pre-explosion reference is the Two Micron All Sky Survey (2MASS), which is shallow and has a very broad PSF.  We instead used an SDSS $z$-band image, but adjusted the flux scale visually to ensure that the extended features of the host galaxy are removed.  
Photometry was performed uniformly on the subtracted images using a custom IDL-based aperture photometry tool.  Calibration of the field was established by comparison of stars in unsubtracted images to SDSS (or, for NIR images, to 2MASS).  SDSS $ugriz$ magnitudes of calibration stars are transformed via the Lupton equations to $BVRI$.   

The transient is very blue compared to any other object in the field: for example, the transient $u-g$ colour is typically $\sim -0.4$ for most of its evolution, compared to a range between +1.48 and +3.04 for bright stars within 5$\arcmin$.  This greatly magnifies the impact of small differences between filter transmission curves for different telescopes (and other wavelength-dependent transmission differences), leading to offsets between different instruments. 

Colour terms for the LT optical filters have been determined by \cite{Smith+2017}.  We colour-corrected SDSS reference stars in the field to the LT system, setting the zeropoint of the transformation as appropriate for an AB colour of 0.0 in all filters.  We then re-calculated the magnitudes of a series of SDSS bright reference star magnitudes using a set of LT exposures taken under the best weather conditions, and used these as secondary standards for the photometry of all LT images (we employ aperture photometry via a custom routine and seeing-matched apertures.)  An additional minor adjustment was made to the $B$ filter ($-0.05$ mag) to match our spectrophotometry (\S \ref{sec:spectra}).   For all other telescopes, we calibrated directly to the SDSS magnitudes, but applied an additional, filter-specific constant adjustment to align each filter to the interpolated LT curve in the same filter and remove any systematic offset.

\begin{figure*}
 \includegraphics[width=6.5in]{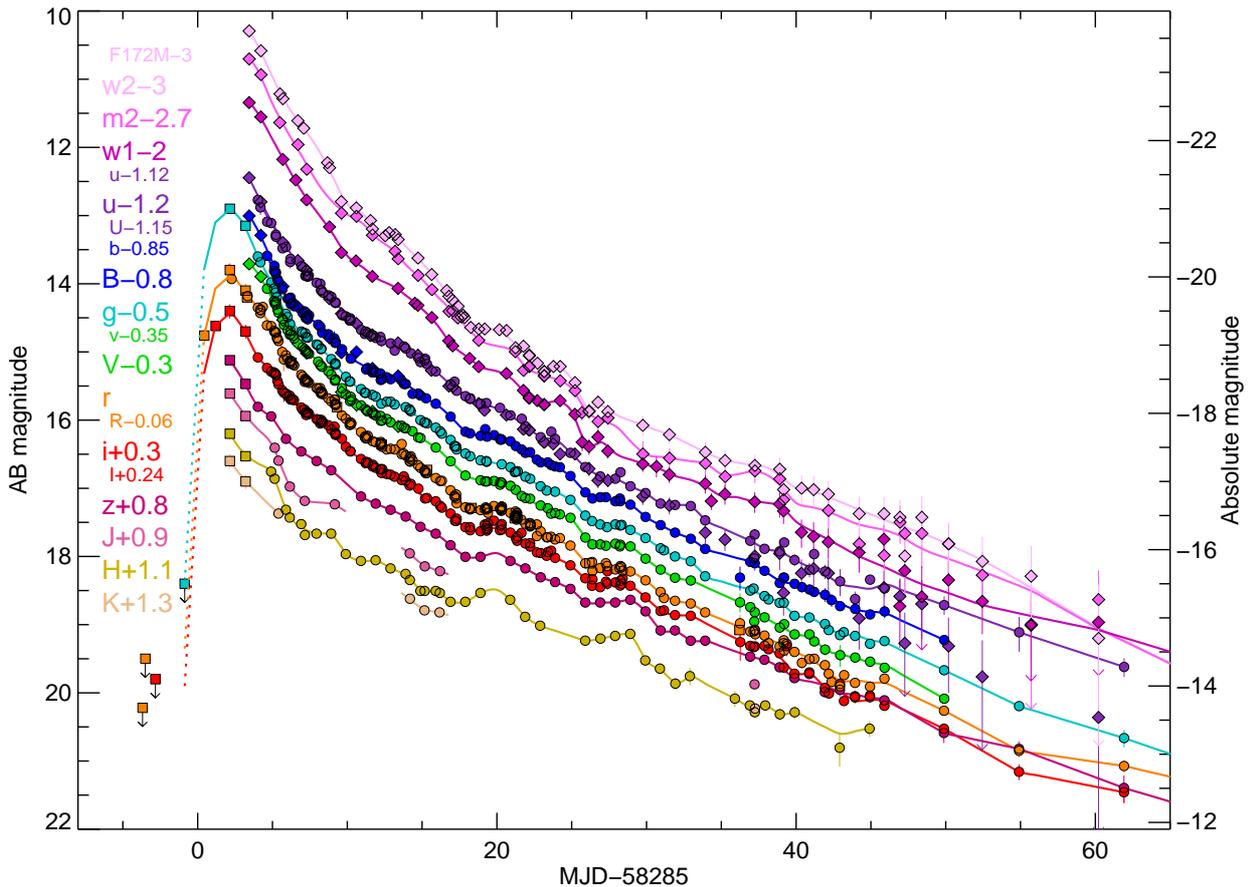}
    \caption{Multi-band light curves of the ultraviolet, optical, and near-infrared transient AT2018cow.  Small offsets have been applied to the filters for clarity (shown at left).  The offsets for the $R$, $I$, and $U$ bands, and of the Swift optical filters, have been chosen to align them with the closest optical bands.  Only the earliest ZTF and ATLAS observations show a rise: from the first epoch of follow-up the transient fades monotonically with time and experiences no subsequent rise in any band, except for short-lived 1--2 day flares in the near-IR.  The coloured curves show a non-parametric interpolation of the observed points in each filter.  The line segments on the rise show a simple linear interpolation or the early transient based on available ATLAS, ZTF, and ASAS-SN data assuming no colour evolution.  Circles show our ground-based data, diamonds show space-based data, and squares indicate photometric measurements from the literature.  Arrows on error bars indicate marginal ($<2\sigma$) UVOT detections.}
    \label{fig:lc}
\end{figure*}

A subset of our photometry is presented in Table \ref{tab:photometry}, and the light curves are plotted in Figure \ref{fig:lc}.

% Photometry table
\begin{table}
    \caption{Early photometric observations of AT\,2018cow from our campaign.  No correction for Galactic extinction has been applied.  A machine-readable table of all 949 photometric data points will be made available online.}
    \centering
    \label{tab:photometry}
   \begin{tabular}{lccll} 
\hline
     MJD & Instrument & Filter & AB magnitude \\
\hline
58287.2674 & P60/SEDM    & r    &  13.93 $\pm$  0.03\\
58288.3405 & P60/SEDM    & r    &  14.18 $\pm$  0.03\\
58288.4416 & Swift/UVOT  & w1   &  13.34 $\pm$  0.05\\
58288.4421 & Swift/UVOT  & u    &  13.57 $\pm$  0.05\\
58288.4426 & Swift/UVOT  & b    &  13.85 $\pm$  0.04\\
58288.4442 & Swift/UVOT  & w2   &  13.29 $\pm$  0.06\\
58288.4448 & Swift/UVOT  & v    &  14.06 $\pm$  0.05\\
58288.4464 & Swift/UVOT  & m2   &  13.40 $\pm$  0.05\\
58289.0227 & LT/IO:O     & u    &  13.97 $\pm$  0.03\\
58289.0234 & LT/IO:O     & g    &  14.10 $\pm$  0.03\\
58289.0241 & LT/IO:O     & r    &  14.35 $\pm$  0.03\\
58289.0248 & LT/IO:O     & i    &  14.78 $\pm$  0.03\\
58289.0255 & LT/IO:O     & z    &  15.01 $\pm$  0.03\\
58289.1889 & KP84/KPED   & g    &  14.18 $\pm$  0.03\\
58289.1901 & KP84/KPED   & r    &  14.43 $\pm$  0.04\\
58289.1904 & P60/SEDM    & r    &  14.38 $\pm$  0.03\\
58289.1963 & KP84/KPED   & U    &  14.03 $\pm$  0.10\\
58289.2108 & P60/SEDM    & r    &  14.39 $\pm$  0.03\\
58289.2229 & Swift/UVOT  & w1   &  13.55 $\pm$  0.03\\
58289.2246 & Swift/UVOT  & u    &  13.92 $\pm$  0.05\\
58289.2263 & Swift/UVOT  & b    &  14.14 $\pm$  0.04\\
58289.2281 & Swift/UVOT  & w2   &  13.58 $\pm$  0.03\\
58289.2298 & Swift/UVOT  & v    &  14.23 $\pm$  0.04\\
58289.2331 & Swift/UVOT  & m2   &  13.63 $\pm$  0.05\\
58289.3493 & P60/SEDM    & r    &  14.34 $\pm$  0.03\\
58289.6299 & HCT/HFOSC   & R    &  14.67 $\pm$  0.03\\
58289.6336 & HCT/HFOSC   & I    &  15.00 $\pm$  0.03\\
58289.6365 & HCT/HFOSC   & V    &  14.37 $\pm$  0.03\\
58289.6397 & HCT/HFOSC   & B    &  14.39 $\pm$  0.03\\
58289.6434 & HCT/HFOSC   & U    &  14.24 $\pm$  0.03\\
58289.9081 & LT/IO:I     & H    &  15.66 $\pm$  0.03\\
58289.9131 & LT/IO:O     & z    &  15.15 $\pm$  0.03\\
58289.9136 & LT/IO:O     & i    &  14.99 $\pm$  0.03\\
58289.9142 & LT/IO:O     & r    &  14.62 $\pm$  0.03\\
58289.9147 & LT/IO:O     & g    &  14.48 $\pm$  0.03\\
58289.9154 & LT/IO:O     & u    &  14.31 $\pm$  0.03\\
%58290.1050 & LT/IO:O     & B    &  14.61 $\pm$  0.03\\
%58290.1061 & LT/IO:O     & V    &  14.55 $\pm$  0.03\\
%58290.1071 & LT/IO:O     & r    &  14.69 $\pm$  0.03\\
%58290.1082 & LT/IO:O     & g    &  14.57 $\pm$  0.03\\
%58290.1093 & LT/IO:O     & i    &  15.01 $\pm$  0.03\\
\hline
\end{tabular}
\end{table}

% Light curve figure

\subsection{Swift Observations}

Observations of AT2018cow using the Neil Gehrels Swift Observatory (\emph{Swift}; \citealt{Gehrels+2004}) began at MJD 58288.442.  Data were collected with both the Ultraviolet-Optical Telescope (UVOT; \citealt{Roming+2005}) and the X-ray Telescope (XRT; \citealt{Burrows+2005}).  The transient was well-detected in both instruments (e.g., \citealt{ATEL11737}) and remained so for the entire monitoring period discussed in this paper.  

Raw UVOT images were processed by the pipeline provided by the {\it Swift} Data Center at the Goddard Space Flight Center (GSFC). The reduced level 2 sky images were downloaded for photometry.  We used the software package {\tt uvotsource} and an aperture radius of $3\arcsec$, chosen to minimize the contamination from the extended host galaxy.  The final photometry output from {\tt uvotsource} was corrected for aperture loss using the curve-of-growth method.

The background was computed from an off-target sky region without any other sources using an aperture radius of $10\arcsec$. The image frames were visually inspected and frames with large pointing smearing were thrown away. For a small number of frames with slight PSF smearing, we used an aperture radius of $5\arcsec$. For frames with astrometric errors, we manually provide the correct centroids as the input to {\tt uvotsource}.

As the UVOT PSF is stable, we subtracted off the estimated host galaxy contribution to the UVOT PSF in flux space rather than via image subtraction.  Photometry from a final epoch (acquired 120.45 days after the reference epoch) was used to estimate the magnitudes within our aperture.  In principle, this final epoch could have contained a small amount of transient flux, although the fact that the optical bands are fading steeply between 50--80 days while negligible fading is seen in the UVOT between 60--120 days suggest that this contribution is very small.)

The XRT data were analysed using an automated reduction routine following the techniques of \citealt{Butler+2007} and binned to increase the S/N.  We assume negligible host contamination (although we note that the galaxy likely hosts a weak AGN; \S \ref{sec:environment}).

\subsection{Astrosat Observations}

AT2018cow was observed by the UltraViolet Imaging Telescope (UVIT; \citealt{Kumar+2012}) on-board AstroSat on 2018-07-03 from 13:45:58 UT to 19:54:12 UT (ToO). These observations were performed in the FUV F172M filter with a total exposure time of 5667 seconds. Images were pre-processed with UVIT L2 pipeline. Aperture photometry was performed using IRAF using an 18-pixel (7.5$\arcsec$) aperture, and calibrated following the calibration procedure mentioned in \cite{Tandon+2017}.

\subsection{Other Photometry}

In addition to our own photometry we also acquire data from public sources and the literature.  In particular, we use the first two epochs of GROND observations from \cite{Prentice+2018} to extend our multicolour optical-NIR coverage to earlier times: we caution that these observations are not host-subtracted or colour-corrected and the aperture size is unknown, although the transient was extremely bright at this time ($\sim14$ mag) and the host contribution should be negligible.  We also use the first epoch of ATLAS photometry from \cite{Prentice+2018}, $r$-band data from the Palomar 48-inch telescope taken as part of the public ZTF Northern Sky Survey, the ZTF $i$-band point reported by \cite{ATEL11738}, and the ASAS-SN limit from \cite{Prentice+2018}.  As these come from imaging-differenced surveys, no host correction is necessary.

\subsection{Optical and Near-IR Spectroscopy}
\label{sec:spectra}

We conducted an extensive campaign to spectroscopically monitor the evolution of the transient at high cadence.  Spectroscopic observations began at MJD 58287.268 (1.82 days after the first ATLAS detection, making this the earliest spectrum obtained of the transient that has been reported so far), and continued at least nightly and usually 2--3 times nightly during the first 12 days after peak.  Sub-night cadence during this period was enabled by observations using spectrographs in California, the Canary Islands, and India: specifically, the SED Machine (SEDM) on the Palomar 60-inch Telescope \citep{Blagorodnova+2018}, the Spectrograph for the Rapid Acquisition of Transients (SPRAT; \citealt{Piascik+2014}) on the Liverpool Telescope, and the Himalayan Faint Object Spectrograph Camera (HFOSC) on the the Himalayan Chandra Telescope.  

\begin{figure}
\includegraphics[width=\columnwidth]{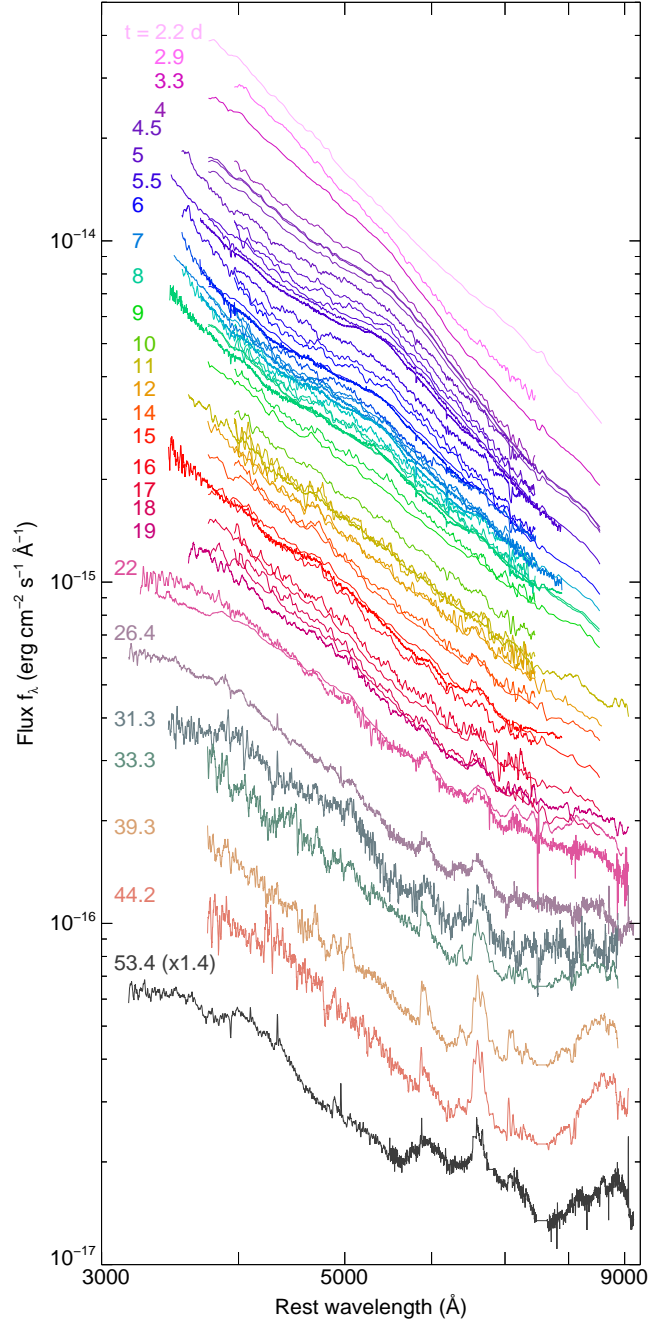}
    \caption{Our full sequence of spectroscopic observations of AT2018cow.  Numbers indicate the time in days since MJD 58285; between days 4--22 they indicate approximate times.  No scaling has been applied: the relative offsets are due to the intrinsic, steady fading of the source.  (The $t=31.3$d and $t=$53.4d spectra have been slightly scaled for clarity.)   We interpolate over host narrow features and (when not corrected) over the telluric A+B bands.  Obvious spectral features develop only at late times, although a very broad, blue dip is visible in all spectra between 4--8 days post-explosion.}
    \label{fig:specsequence}
\end{figure}

Additional spectra were obtained less regularly and at later phases using larger telescopes: the DeVeny spectrograph at the Discovery Channel Telescope (DCT), the Andalucia Faint Object Spectrograph and Camera (ALFOSC) on the Nordic Optical Telescope (NOT), the Double-Beam Spectrograph (DBSP; \citealt{Oke+1982}) and the TripleSpec near-infrared spectrograph on the 200-inch Hale Telescope, the Gemini Multi-Object Spectrograph (GMOS) on Gemini-North, and the Low-Resolution Imaging Spectrograph (LRIS; \citealt{Oke+1995}) at Keck Observatory. 
A log of all spectroscopic observations can be found in Table \ref{tab:spectroscopy}, and all spectra are plotted in Figure \ref{fig:specsequence}.

% Spectroscopy log table
\begin{table}
    \caption{Log of spectroscopic observations of AT\,2018cow.  Times are relative to the reference epoch of MJD 58285.}
    \label{tab:spectroscopy}
    \centering
    \begin{tabular}{lccll}
        \hline
        MJD & $t$ (d) & Exp. (s) & Telescope & Instrument \\
        \hline
58287.268 &  2.268 &  1600 & P60     & SEDM     \\
58287.949 &  2.949 &   300 & LT      & SPRAT    \\
58288.341 &  3.341 &  1600 & P60     & SEDM     \\
58289.000 &  4.000 &   180 & LT      & SPRAT    \\
58289.191 &  4.191 &  1600 & P60     & SEDM     \\
58289.211 &  4.211 &  1600 & P60     & SEDM     \\
58289.350 &  4.350 &  1600 & P60     & SEDM     \\
58289.651 &  4.651 &   900 & HCT     & HFOSC    \\
58289.946 &  4.946 &   180 & LT      & SPRAT    \\
58290.097 &  5.097 &   450 & LT      & SPRAT    \\
58290.196 &  5.196 &  2500 & P60     & SEDM     \\
58290.261 &  5.261 &   250 & DCT     & DeVeny   \\
58290.353 &  5.353 &   300 & P200    & DBSP     \\
58290.327 &  5.327 &  1800 & Gemini-N &    GMOS \\ 
58290.618 &  5.618 &  1200 & HCT     & HFOSC    \\
58291.020 &  6.020 &   450 & LT      & SPRAT    \\
58291.224 &  6.224 &  2500 & P60     & SEDM     \\
58291.276 &  6.276 &  4800 & P200    & TripleSpec \\
58291.337 &  6.337 &  1800 & Gemini-N &    GMOS \\ 
58291.636 &  6.636 &  1000 & HCT     & HFOSC    \\
58291.939 &  6.939 &   240 & LT      & SPRAT    \\
58292.027 &  7.027 &   450 & LT      & SPRAT    \\
58292.145 &  7.145 &   180 & DCT     & DeVeny   \\
58292.181 &  7.181 &  2500 & P60     & SEDM     \\
58292.374 &  7.374 &  1800 & Gemini-N   &  GMOS \\ 
58292.648 &  7.648 &  1200 & HCT     & HFOSC    \\
58292.955 &  7.955 &   300 & LT      & SPRAT    \\
58293.018 &  8.018 &   450 & LT      & SPRAT    \\
58293.182 &  8.182 &  2500 & P60     & SEDM     \\
58293.212 &  8.212 &  2500 & P60     & SEDM     \\
58293.288 &  8.288 &  1800 & Gemini-N   &  GMOS \\ 
58293.821 &  8.821 &  1200 & HCT     & HFOSC    \\
58293.892 &  8.892 &   300 & LT      & SPRAT    \\
58294.182 &  9.182 &  2500 & P60     & SEDM     \\
58294.656 &  9.656 &  1200 & HCT     & HFOSC    \\
58294.989 &  9.989 &   300 & LT      & SPRAT    \\
58295.894 & 10.894 &   240 & LT      & SPRAT    \\
58296.017 & 11.017 &   600 & NOT     & ALFOSC   \\
58296.103 & 11.103 &   450 & LT      & SPRAT    \\
58296.913 & 11.913 &   240 & LT      & SPRAT    \\
58297.245 & 12.245 &  2500 & P60     & SEDM     \\
58297.349 & 12.349 &  1800 & P200    & TripleSpec \\
% 58298.722 & 13.722 &  1800 & HCT     & HFOSC    \\  excluded
58298.916 & 13.916 &   240 & LT      & SPRAT    \\
58299.212 & 14.212 &  2500 & P60     & SEDM     \\
58299.766 & 14.767 &  2400 & HCT     & HFOSC    \\
58300.180 & 15.180 &  2500 & P60     & SEDM     \\
58300.389 & 15.389 &   900 & Gemini-N   &  GMOS \\ 
58300.622 & 15.622 &  2400 & HCT     & HFOSC    \\
58300.896 & 15.896 &   240 & LT      & SPRAT    \\
58301.990 & 16.990 &   600 & LT      & SPRAT    \\
58302.275 & 17.275 &  2500 & P60     & SEDM     \\
58302.908 & 17.908 &   360 & LT      & SPRAT    \\
58303.180 & 18.180 &  2500 & P60     & SEDM     \\
58304.000 & 19.028 &   900 & NOT     & ALFOSC   \\
58307.034 & 22.034 &   900 & NOT     & ALFOSC   \\
58307.301 & 22.301 &  1200 & P200    & DBSP     \\
58311.397 & 26.397 &  1800 & Keck I  & LRIS     \\
58316.345 & 31.345 &   600 & P200    & DBSP     \\
58318.295 & 33.295 &  1200 & Gemini-N & GMOS    \\
58324.300 & 39.300 &  1800 & Gemini-N & GMOS    \\
58329.254 & 44.254 &  1800 & Gemini-N & GMOS    \\
58338.359 & 53.359 &  3180 & Keck I   & LRIS    \\
\hline
    \end{tabular}
\end{table}

LT/SPRAT and P60/SEDM data were processed by automated reduction pipelines designed for each facility\footnote{The SEDM pipeline is described at http://www.astro.caltech.edu/sedm/Pipeline.html; the SPRAT pipeline is a modification of the pipeline for FrodoSpec \citep{Barnsley+2012}}. The LPipe reduction pipeline\footnote{http://www.astro.caltech.edu/~dperley/programs/lpipe.html} (Perley et al. 2018, in prep) was used to process the LRIS data.  Reductions for the remaining spectrographs were performed manually using standard IRAF tools.

After initial reduction and flux calibration, all spectra were absolutely calibrated by comparing synthetic photometry of the spectrum to photometry from our imaging data.  The absolute flux scale is established by comparing synthetic $r$-band photometry calculated from each spectrum to our (true) $r$-band photometry, interpolated to the appropriate epoch.  To correct for imperfections in the calibration related to atmospheric attenuation or wavelength-dependent slit losses, we next colour-correct the spectrum by comparing a synthetic $g-r$ colour to the true photometric $g-r$ colour, and warping the spectra by a power-law correction factor.\footnote{The colour correction was typically quite small: $<$0.1 mag in nearly all cases.}  Since the spectra unavoidably include some host-galaxy light, we re-add an estimate of the host galaxy flux within the slit to the photometry (estimated given the size of the slit and using our host-galaxy model; \S \ref{sec:hostgalaxy}) prior to the photometric correction, and subtract the host galaxy model after correction.

\section{Observational Properties}
\label{sec:properties}

\begin{table}
    \centering
    \caption{Key properties of AT\,2018cow}
    \label{tab:properties}
    \begin{tabular}{lll} % four columns, alignment for each
        \hline
$z $  & 0.0140 & Redshift (from host emission) \\
$t_{\rm rise} $  & $\sim$2.5 d  & Rise time to peak ($g$)\\
$t_{{\rm rise},1/2} $  & $\sim$1.5 d & Time to rise from half-max ($r$) \\
$t_{{\rm decline},1/2} $ & $\sim$3 d & Time to decay to half-max ($r$) \\
$M_{g,{\rm peak}}$ & $-$20.4 & Peak $g$ absolute magnitude \\
$M_{r,{\rm peak}}$ & $-$19.9 & Peak $r$ absolute magnitude \\
$L_{\rm bol, peak}$ & 4$\times$10$^{44}$ erg s$^{-1}$& UVOIR luminosity at optical peak\\
$T_{\rm char}$    & 17000 K & Characteristic temperature\\
$E_{\rm rad}$     & 5$\times$10$^{49}$ erg & Total UVOIR radiative output \\
$v_{\rm spec}$    & 6000 km s$^{-1}$ & Velocity width of late emission lines \\
$M_{\rm *,host}$  & 1.4$\times$10$^{9}$ $M_\odot$ & Host stellar mass \\
SFR$_{\rm host}$  & 0.22 $M_\odot$yr$^{-1}$ & Host star-formation rate \\
        \hline
\end{tabular}
\end{table}

\subsection{Environment and Pre-Explosion Constraints}
\label{sec:environment}

The transient lies on the sky coincident with the catalogued galaxy CGCG 137-068, an unremarkable dwarf spiral galaxy showing a faint bar and spiral arms (Figure \ref{fig:image}).  Two sources are present within the SDSS and PS1 pre-imaging: a reddish point source at the galaxy nucleus (likely a weak AGN) and a compact, but not truly pointlike, source approximately 1.9$\arcsec$ east-southeast of the transient (probably an HII region).  AT\,2018cow is located far from the centre of the galaxy (5.9$\arcsec$ or 1.7 kpc from the nucleus), and no point or pointlike source is visible at the location of the transient itself.  Forced photometry on a median filtered PS1 image limits any contribution from an unresolved source to $g>22.2$, $r>22.3$, $i>21.9$: more than 8 magnitudes below the transient at peak.

Additionally, we checked for evidence of pre-explosion variability in both the Catalina Real-Time Survey and iPTF archives.  We found no evidence for any previous outbursts from the location of the transient.

\subsection{A Fast, Consistently Blue Transient}

Light curves of the transient, assembled by our worldwide telescope network, are shown in Figure \ref{fig:lc}.  The photometric properties alone exhibit several remarkable features unprecedented for any other extragalactic transient observed at this level of detail.

As we have already noted, the rise time is very fast.  Comparing the ATLAS $o$ discovery magnitude (which is dominated by $r$ flux for this blue transient) to the GROND $r$ magnitude indicates a rise from half-max of only 1--2 days.  The ASAS-SN $g-$band limit suggests an explosion time of no more than 1 day prior to the discovery observation, giving a total time from explosion to peak of between 2--3 days.

The transient is extremely luminous at peak ($M_r \sim -19.9$ or $M_g \sim -20.4$).  This is more luminous than any core-collapse supernova with the exception of a small fraction of Type IIn and superluminous supernovae, both of which exhibit very long rise and decay times.

The fading, like the rise, is quite rapid.  The time to decline to half of its peak flux is only about 4 days, and there is no subsequent rise to a second, radioactively powered peak: the light curve fades monotonically (except in the NIR, which exhibits minor but significant fluctuations on timescales of 2--3 days).  By around 25 days post-discovery the transient has a luminosity ($M_r \sim -16$) well below that of a typical core-collapse supernova at the same phase.  

Finally, the colour is extremely and persistently blue.  In early observations the colours are close to the Rayleigh-Jeans power-law limit, indicating a thermal origin with a spectral peak far into the UV (\S \ref{sec:physevol}).   Hot, blue early phases of supernovae are common (shortly after shock breakout and before adiabatic losses have cooled the ejecta), but AT\,2018cow retains a high temperature for a remarkably long period: after a month, the optical colours are bluer than most SNe are even in their earliest phases and it remains well-detected in all UV filters.

These properties are summarized in Table \ref{tab:properties}.  \cite{Prentice+2018} also independently report the exceptionally fast evolution and blue color of this transient, as does the recent analysis by \cite{Margutti+2018}.

\subsection{Spectral evolution}

\begin{figure*}
\includegraphics[width=6.5in]{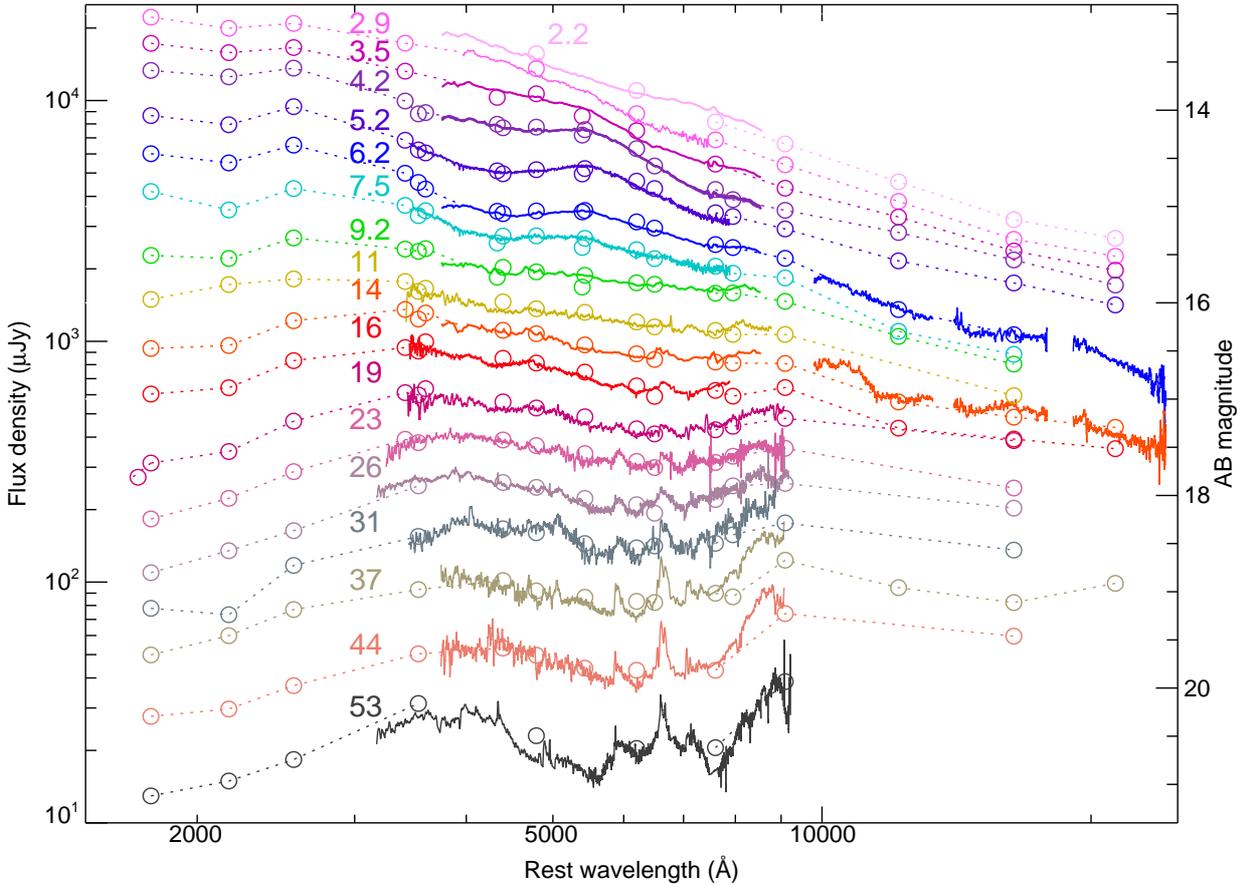}
    \caption{Spectral energy distribution (in $f_\nu$) sequence of AT2018cow from the UV to the NIR, with selected spectra overplotted.  (The closest high-quality spectrum to each photometric reference epoch is shown, rescaled by a constant factor to match the absolute flux level).  The spectrum is initially (days 2--3) hot and featureless.  A broad absorption feature develops in the UV/blue region of the spectrum starting around day 4, but disappears again by day 9.  Narrower features begin to emerge after $>10$ days, and the NIR bands become dominated by a red SED component that peaks around 10000\,\AA.  Our photometry and spectroscopy show good consistency (except in $z$-band at late times).  In particular, both show the early, broad spectral feature between 3500--5500\,\AA.}
    \label{fig:sedsequence}
\end{figure*}

The behaviour seen in the spectra is also unprecedented.  The earliest spectra in our sequence (Figure \ref{fig:sedsequence}), sampling close to the peak time of the transient, show only a hot and smooth continuum: they are particularly lacking in emission or absorption features, except for weak emission from host galaxy H$\alpha$ (not shown in our figures since we interpolate over the host narrow lines).  There is no sign of any flash-ionized emission features \citep[e.g.,][]{GalYam+2014,Yaron+2017,Khazov+2016}.

Beginning around MJD 58299 (day 4 on our plots), a single, extremely broad feature begins to emerge in all of our spectra and in our photometry.  If interpreted as an absorption trough, its centre is at approximately 4600\,\AA\ with a full-width of 1500\,\AA.  It is vaguely reminiscent of the Fe II feature seen in broad-lined Ic supernovae around peak light (e.g. \citealt{Galama+1998}), a resemblance that led to early suggestions of a Ic-BL classification \citep{ATEL11740,ATEL11753}.  Simultaneously with the emergence of this feature, a very bright radio/submillimeter afterglow was detected \citep{ATEL11749,ATEL11774,ATEL11795} which---at the time---seemed to seal the Ic-BL association and led to anticipation that these features would strengthen and a supernova peak would emerge shortly in the light curve.  

This is not what happened: while the feature strengthens slightly between days 4 and 5, from then on it begins to dissipate and by day 8 it vanishes entirely, returning to a largely featureless blue continuum \citep{ATEL11776}.

\begin{figure}
\includegraphics[width=\columnwidth]{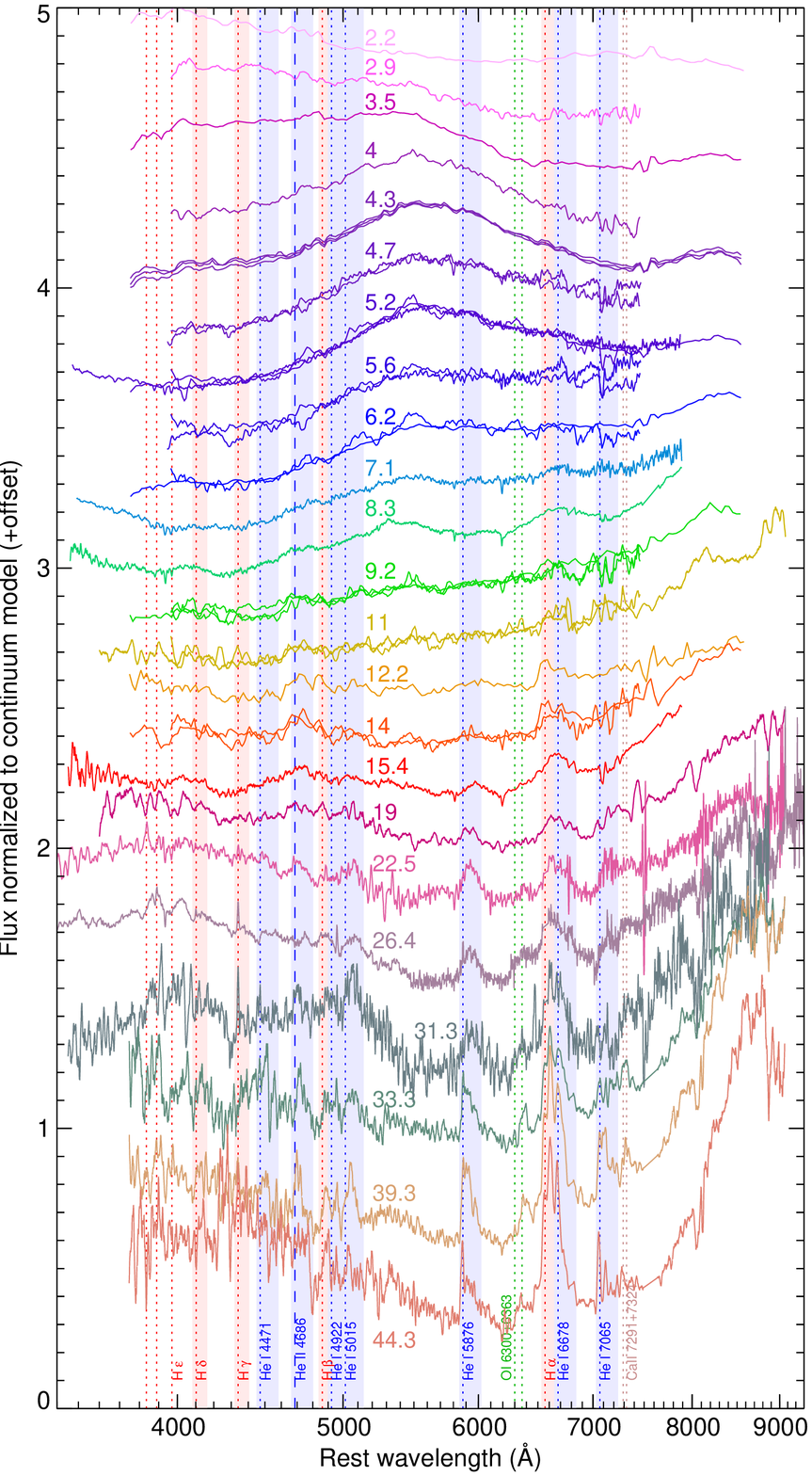}
    \caption{Sequence showing a subset of our spectra after division and normalization by a blackbody model, fit to the coeval photometry (Figure \ref{fig:bbfits}).  Line identifications are shown as vertical coloured bars, all of which emerge only at later times.  Thin dashed lines show the rest wavelength of each transition, while the shaded bands show the approximate observed widths of the emission component.}
    \label{fig:normspectra}
\end{figure}

\begin{figure}
\includegraphics[width=\columnwidth]{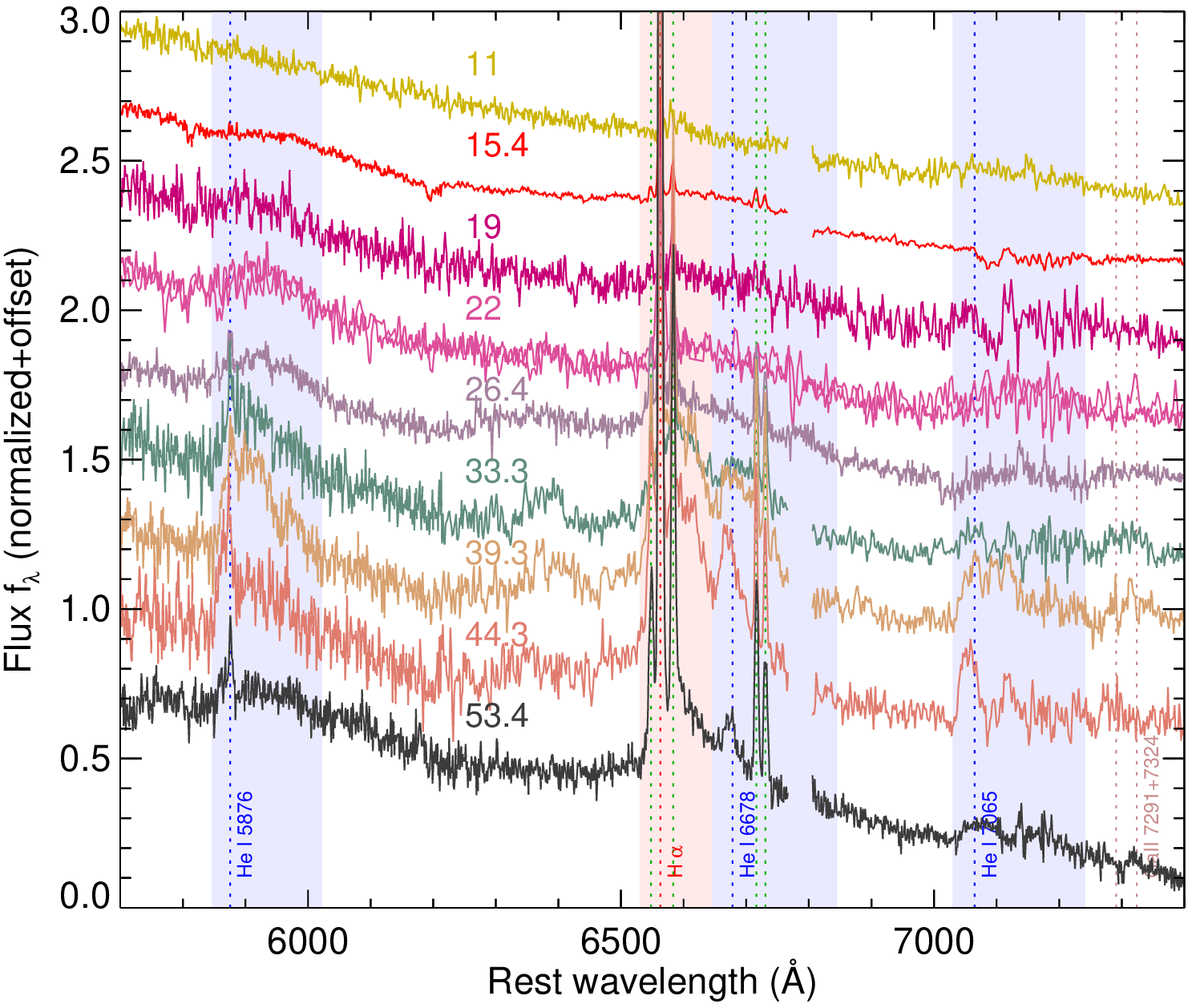}
    \caption{Late-time spectra of AT\,2018cow showing the central region of the spectrograph.  Host-galaxy emission has not been removed; the gap is the telluric band.  The helium lines are completely absent at +11 days, but begin to appear at +15 days.  At +30 days they develop a weak blueshifted narrow component.}
    \label{fig:latespec}
\end{figure}

Very different evolution sets in after this time.  First, a weak and moderately-broad (full-width $\sim200$\AA; v $\sim$ 10000 km/s) emission feature centred at $\approx 4850$\,\AA\ begins to emerge: it is difficult to recognize because spectra during this period are of low quality owing to the presence of the nearly-full moon, but is seen consistently in both the LT and the SEDM spectra on days 9, 11, 12, and 14 (Figure \ref{fig:normspectra}); it was also independently seen in NOT spectra reported by \citep{ATEL11836}.  Its most likely interpretation is \HeII\ $\lambda 4686$.  The line fades thereafter, but a variety of other lines of similar velocity width and offset begin to appear between 20--30 days. Emission features of \HeI\ $\lambda$5876 and \HeI\ $\lambda$5015 are clearly visible starting at $\sim$15 days, along with emission from H$\alpha$ (in a blend with \HeI\ $\lambda$6678), H$\beta$, H$\gamma$, H$\delta$, and a blend of several higher Balmer lines.  All of these lines are significantly and consistently offset to the red by about +3000 km/s at the time of first detection.  However, over the subsequent 10--20 days the profiles evolve blueward, developing a ``wedge'' shape: the peak (which often contains a weak narrow component) is very close to the rest-frame wavelength, with a steep fall towards the blue and a very gradual one towards the red (Figure \ref{fig:latespec}).  Additional lines, including \HeI\ $\lambda$7065, weak \CaII ] $\lambda\lambda$7291,7324, and (possibly) \OI\ $\lambda\lambda$6300, 6363 also begin to emerge at later times ($>30$ days).  A very strong, broad upturn between 8000--9000 (also easily visible in the photometry as a $z$-band excess) emerges around this time as well, although its origin is unclear: its wavelength is close to that of the \CaII\ IR triplet but it is much broader than would be expected from this feature alone if it has a similar profile as the H and He lines, especially on the blue wing.

\subsection{Physical Properties}
\label{sec:physevol}

To characterize the early SED, we first construct coeval sets of photometry by performing a nonparametric interpolation of the light curve for each filter (the same procedure was used in the $g$ and $r$ bands to colour-correct the spectroscopy; Section \ref{sec:spectra}).  Galactic extinction is corrected using the \cite{Fitzpatrick1999} attenuation curve and $E_{B-V}$ = 0.07 \citep{Schlafly2011}.   We assume no extinction in the host galaxy.

The early SEDs are unambiguously thermal.  The UVOIR slope ($F_\nu \propto \nu^{\alpha}$) during the first epoch is $\alpha $=$ 1.2 \pm 0.1$ as measured between the $u$ and the $z$ bands: close to the Rayleigh-Jeans $\alpha$ = 2 and inconsistent (in particular) with synchrotron emission, which exhibits $\alpha $= 0.33 below the peak frequency and $-0.5$ to $-1.25$ above it (e.g., \citealt{Sari+1998}).  The colour of the transient becomes gradually less blue as time passes, but it remains effectively thermal throughout, with the peak (in $\nu F_\nu$) remaining in the UV at all times.  

\begin{figure*}
\includegraphics[width=6.5in]{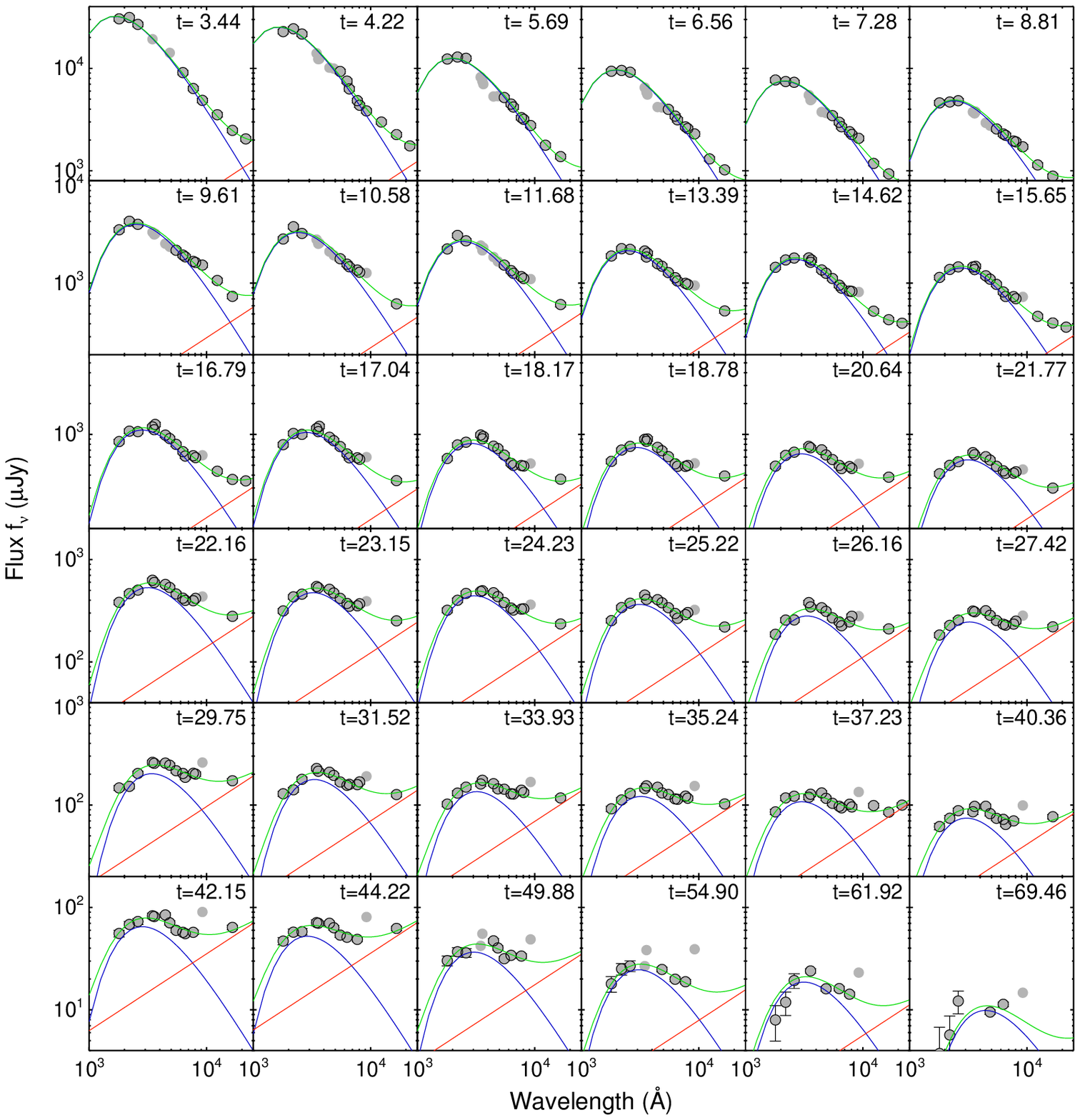}
    \caption{Fits to the multi-epoch photometry using a combined blackbody (blue curve) and power-law (red curve) model.  The green curve shows the sum of these models.  Data points that are not enclosed in circles are excluded from the fits, usually due to contamination by strong emission or absorption features.}
    \label{fig:bbfits}
\end{figure*}

To characterize the evolution of the photosphere, we fit a Planck function to the UV-optical data at the time of each UVOT epoch.  A single Planck function fits the UV and most optical filters well at essentially every epoch, but underpredicts the NIR fluxes after a few days; it also cannot explain the persistent ``dip'' seen in the $uBg$ filters in several early optical observations (Figure \ref{fig:bbfits}).   We thus exclude the $uBg$ filters from the fits, and add an additional red component to the model. The form of this red component is not well-constrained by our data (our light curve coverage in the NIR is very incomplete outside the $H$-band). We tried both a second blackbody and a power-law; we obtain acceptable fits to most bands for a blackbody with a constant, low ($\sim$3000 K) temperature and a power-law with spectral index ($F_\nu \propto \nu^{\alpha}$) of $\alpha \sim -0.75$.  We prefer the power-law model: a warm blackbody is not well-justified theoretically (the observed temperature is too hot to be easily explained as dust, although similar red components have been seen in some SNe; e.g., \citealt{Kangas+2016}), whereas a synchrotron power-law of $\alpha \sim 0.5-1.0$ is expected given the bright radio afterglow (and an extrapolation of the flux to the millimeter band provides reasonable consistency with reported millimeter fluxes).  The $z$-band at late times shows strong excess relative to either model and is excluded from our final fits.  We fix the spectral index at $\alpha=-0.75$ for all epochs.

At very late times ($>45$d) our ground-based coverage becomes sparse, due to both the fading of the transient and the shortening window of observations each night.  At these times we fix our epochs to the ground-based (LT) epochs, interpolating the low-S/N (but numerous) UVOT fluxes via local regression.  We caution that derived parameters in this regime are particularly uncertain due to the absence of NIR coverage, presence of emission features, and systematics associated with the host subtraction.  For the last epoch (65 days) the power-law component could not be constrained and is fixed by extrapolation of the preceding epochs.

\begin{figure}
\includegraphics[width=\columnwidth]{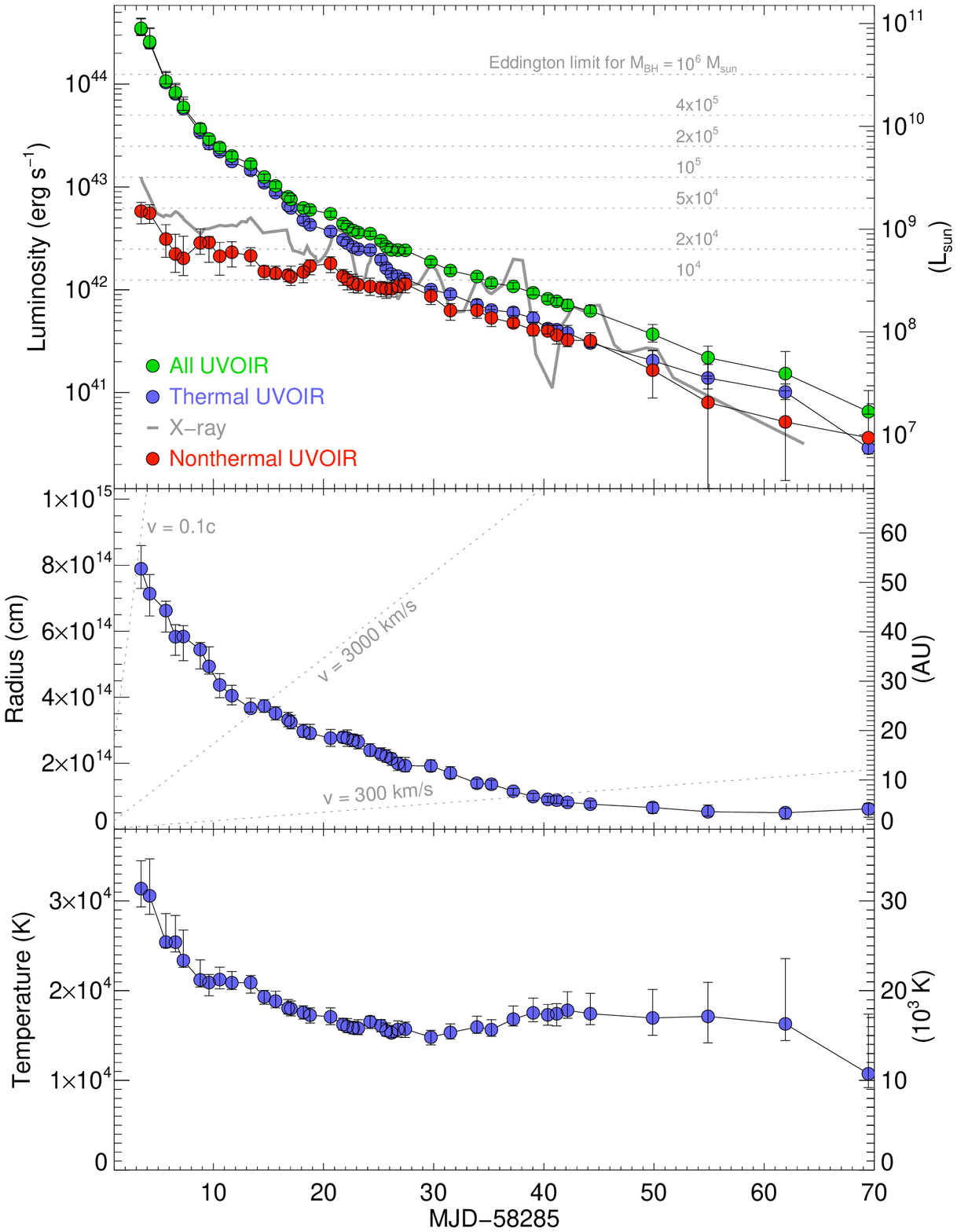}
    \caption{Physical properties based on the blackbody fits shown in Figure \ref{fig:bbfits}.  In the top panel, the green data points show the luminosity inferred by summing the total Stefan-Boltzmann luminosity of the thermal component and an integral of the power-law component for $\lambda>1000$\,\AA.  (Each component is also shown separately as blue and red points, respectively.)  The X-ray light curve is also shown in grey.
}
    \label{fig:physevol}
\end{figure}

Results are plotted in Figure \ref{fig:physevol} and listed in Table \ref{tab:physevol}. At peak, the object is very hot (30000 K) and already quite large in size, with an inferred radius of almost 10$^{15}$\,cm.  This implies fast ejecta: given the ASAS-SN pre-explosion limit, the time of the first SED was only $\sim4$ days after the initial explosion time and thus the expansion speed must exceed $>0.1$c.  Alternatively, the rapid expansion of the photosphere could imply a high-velocity shock traversing pre-existing, optically thick material.  However, the broad absorption feature independently implies that this material must also be traveling at of $>0.1$c at this time, so if the transient is due to an explosion (c.f. \S \ref{sec:tdemodel}) it must represent part of the ejecta.

\begin{table}
    \centering
    \caption{Photospheric parameters derived from a combined synchrotron+blackbody fit to the UV-optical-NIR data.  Uncertainty estimates are statistical errors only.}
    \label{tab:physevol}
    \begin{tabular}{lccll}
        \hline
MJD      &  L (L$_\odot$)     &   R (AU)    &      T (kK)   \\
\hline
58288.44 & 8.96e+10$^{\rm+2.32e+10}_{\rm-1.19e+10}$ &  52.77$^{+4.70}_{-3.98}$ & 31.39$^{+3.10}_{-2.04}$ \\
58289.22 & 6.64e+10$^{\rm+2.46e+10}_{\rm-8.66e+09}$ &  47.72$^{+3.89}_{-4.51}$ & 30.58$^{+4.12}_{-2.06}$ \\
58290.69 & 2.75e+10$^{\rm+6.67e+09}_{\rm-1.70e+09}$ &  44.29$^{+1.89}_{-4.35}$ & 25.42$^{+3.19}_{-0.60}$ \\
58291.56 & 2.13e+10$^{\rm+4.81e+09}_{\rm-2.04e+09}$ &  38.99$^{+2.44}_{-3.74}$ & 25.42$^{+2.98}_{-1.08}$ \\
58292.28 & 1.53e+10$^{\rm+3.91e+09}_{\rm-1.06e+09}$ &  39.04$^{+2.18}_{-4.88}$ & 23.37$^{+3.40}_{-0.77}$ \\
58293.81 & 9.47e+09$^{\rm+1.17e+09}_{\rm-8.92e+08}$ &  36.39$^{+1.45}_{-3.94}$ & 21.20$^{+2.22}_{-0.80}$ \\
58294.61 & 7.53e+09$^{\rm+5.24e+08}_{\rm-1.02e+09}$ &  32.99$^{+3.97}_{-1.54}$ & 20.91$^{+0.90}_{-1.49}$ \\
58295.58 & 6.23e+09$^{\rm+6.60e+08}_{\rm-5.08e+08}$ &  29.24$^{+2.29}_{-2.59}$ & 21.25$^{+1.37}_{-1.02}$ \\
58296.68 & 5.15e+09$^{\rm+5.53e+08}_{\rm-3.42e+08}$ &  27.07$^{+2.10}_{-1.84}$ & 20.90$^{+1.24}_{-0.76}$ \\
58298.39 & 4.29e+09$^{\rm+3.19e+08}_{\rm-4.10e+08}$ &  24.51$^{+2.08}_{-0.78}$ & 20.91$^{+0.78}_{-1.17}$ \\
58299.62 & 3.22e+09$^{\rm+2.12e+08}_{\rm-2.46e+08}$ &  24.96$^{+1.28}_{-1.25}$ & 19.32$^{+0.73}_{-0.83}$ \\
58300.65 & 2.64e+09$^{\rm+2.74e+08}_{\rm-1.64e+08}$ &  23.50$^{+1.31}_{-1.37}$ & 18.84$^{+1.11}_{-0.76}$ \\
58301.79 & 2.06e+09$^{\rm+2.02e+08}_{\rm-9.79e+07}$ &  22.19$^{+1.52}_{-1.40}$ & 18.04$^{+0.97}_{-0.64}$ \\
58302.04 & 1.95e+09$^{\rm+1.64e+08}_{\rm-1.24e+08}$ &  21.69$^{+1.43}_{-1.35}$ & 18.00$^{+0.85}_{-0.80}$ \\
58303.17 & 1.61e+09$^{\rm+1.05e+08}_{\rm-9.91e+07}$ &  19.86$^{+1.42}_{-1.23}$ & 17.57$^{+0.75}_{-0.74}$ \\
58303.78 & 1.54e+09$^{\rm+1.12e+08}_{\rm-1.16e+08}$ &  19.47$^{+1.79}_{-1.20}$ & 17.28$^{+0.81}_{-0.89}$ \\
58305.64 & 1.41e+09$^{\rm+8.66e+07}_{\rm-1.06e+08}$ &  18.46$^{+1.76}_{-1.66}$ & 17.09$^{+1.00}_{-0.87}$ \\
58306.77 & 1.13e+09$^{\rm+8.69e+07}_{\rm-6.57e+07}$ &  18.62$^{+1.13}_{-1.05}$ & 16.23$^{+0.70}_{-0.62}$ \\
58307.16 & 1.06e+09$^{\rm+7.78e+07}_{\rm-7.25e+07}$ &  18.46$^{+1.66}_{-1.64}$ & 16.01$^{+0.87}_{-0.68}$ \\
58307.70 & 9.69e+08$^{\rm+7.34e+07}_{\rm-5.53e+07}$ &  18.07$^{+1.22}_{-1.34}$ & 15.84$^{+0.75}_{-0.66}$ \\
58308.15 & 9.27e+08$^{\rm+8.19e+07}_{\rm-5.71e+07}$ &  17.70$^{+1.39}_{-1.42}$ & 15.81$^{+0.96}_{-0.71}$ \\
58309.23 & 9.04e+08$^{\rm+5.32e+07}_{\rm-5.56e+07}$ &  16.00$^{+1.34}_{-1.22}$ & 16.55$^{+0.62}_{-0.76}$ \\
58310.22 & 7.73e+08$^{\rm+4.16e+07}_{\rm-5.47e+07}$ &  15.19$^{+1.31}_{-0.78}$ & 16.10$^{+0.68}_{-0.69}$ \\
58310.70 & 6.79e+08$^{\rm+5.02e+07}_{\rm-4.02e+07}$ &  14.78$^{+1.37}_{-1.18}$ & 15.55$^{+0.84}_{-0.60}$ \\
58311.16 & 6.29e+08$^{\rm+5.06e+07}_{\rm-3.89e+07}$ &  14.30$^{+1.18}_{-1.39}$ & 15.30$^{+0.83}_{-0.55}$ \\
58311.76 & 6.26e+08$^{\rm+4.82e+07}_{\rm-3.94e+07}$ &  13.33$^{+1.28}_{-1.45}$ & 15.67$^{+0.98}_{-0.84}$ \\
58312.42 & 6.23e+08$^{\rm+4.23e+07}_{\rm-5.01e+07}$ &  12.88$^{+1.67}_{-1.21}$ & 15.71$^{+0.81}_{-0.93}$ \\
58314.75 & 4.82e+08$^{\rm+2.74e+07}_{\rm-3.48e+07}$ &  12.82$^{+1.25}_{-1.10}$ & 14.82$^{+0.76}_{-0.86}$ \\
58316.52 & 3.96e+08$^{\rm+2.91e+07}_{\rm-2.75e+07}$ &  11.38$^{+1.32}_{-1.11}$ & 15.34$^{+0.95}_{-0.73}$ \\
58318.93 & 3.46e+08$^{\rm+2.72e+07}_{\rm-2.22e+07}$ &   9.34$^{+0.84}_{-1.00}$ & 15.94$^{+1.22}_{-0.70}$ \\
58320.24 & 3.00e+08$^{\rm+2.92e+07}_{\rm-2.24e+07}$ &   9.13$^{+0.85}_{-0.96}$ & 15.66$^{+1.12}_{-0.74}$ \\
58322.23 & 2.78e+08$^{\rm+2.42e+07}_{\rm-1.63e+07}$ &   7.70$^{+0.57}_{-0.80}$ & 16.83$^{+1.46}_{-0.76}$ \\
58324.03 & 2.41e+08$^{\rm+2.54e+07}_{\rm-1.14e+07}$ &   6.64$^{+0.65}_{-0.79}$ & 17.53$^{+1.64}_{-0.98}$ \\
58325.36 & 2.10e+08$^{\rm+1.94e+07}_{\rm-1.20e+07}$ &   6.06$^{+0.60}_{-0.64}$ & 17.31$^{+1.16}_{-1.15}$ \\
58326.17 & 1.99e+08$^{\rm+1.70e+07}_{\rm-1.87e+07}$ &   5.91$^{+0.88}_{-0.69}$ & 17.43$^{+1.13}_{-1.37}$ \\
58327.15 & 1.81e+08$^{\rm+2.66e+07}_{\rm-1.22e+07}$ &   5.46$^{+0.45}_{-0.72}$ & 17.81$^{+2.07}_{-0.88}$ \\
58329.22 & 1.60e+08$^{\rm+2.16e+07}_{\rm-8.68e+06}$ &   5.09$^{+0.64}_{-0.75}$ & 17.43$^{+2.27}_{-1.21}$ \\
58334.88 & 9.52e+07$^{\rm+2.32e+07}_{\rm-1.52e+07}$ &   4.41$^{+1.01}_{-1.16}$ & 16.96$^{+3.17}_{-1.93}$ \\
58339.90 & 5.63e+07$^{\rm+1.67e+07}_{\rm-1.90e+07}$ &   3.56$^{+1.38}_{-0.86}$ & 17.14$^{+3.79}_{-2.95}$ \\
58346.92 & 3.94e+07$^{\rm+2.52e+07}_{\rm-8.20e+06}$ &   3.37$^{+0.90}_{-1.33}$ & 16.30$^{+7.28}_{-1.86}$ \\
58354.46 & 1.68e+07$^{\rm+1.02e+07}_{\rm-9.16e+05}$ &   4.14$^{+1.09}_{-1.70}$ & 10.74$^{+6.65}_{-1.54}$ \\
        \hline
\end{tabular}
\end{table}

Surprisingly, after this initial rise no further expansion is inferred: the photospheric radius \emph{declines} continuously throughout our observations.  This is extremely unusual for a supernova: normally, the photosphere expands with the expanding material in the early, optically-thick phases.  

The temperature initially declines with time, as expected for most explosive transients.   However this parameter, too, begins to exhibit unusual evolution at later epochs: after 20 days the temperature curve levels off and in subsequent epochs it actually \emph{increases}, levelling off at about 17000 K before possibly falling again in the last epoch.  The significance of the late increase is dependent on the SED model (and in particular the treatment of the red excess) and on the UV host subtraction procedure, but the temperature is, in any case, still extremely high 1--2 months after peak light.

The bolometric luminosity of the transient decays in a remarkably simple fashion similar to a power-law in time.  Setting $t$=$0$ to our reference epoch of MJD 58285, the temporal index ($F \propto t^{\beta}$) is $\beta \sim -2.5$, steeper than but not remarkably different from the classical -5/3 expected for TDEs and similar accretion-powered events.

We have plotted the luminosity of the two fitted components (the thermal peak and the possibly non-thermal red component) separately in the top panel of Figure \ref{fig:physevol}; the non-thermal component is integrated only at $\lambda>1000$\,\AA.  The non-thermal flux shows a similar average decay as the X-ray (supporting the notion that it arises from a physically distinct region from the thermal emission) but does not show the same strong temporal variations (see also \citealt{Rivera+2018}), so it is not clear whether they truly represent the same component.  However, the red bands do show much greater variability than the bluer filters at early times: this is best illustrated by an apparent $i$-band bump at 20 days visible in Figure \ref{fig:lc}. (Unfortunately, this event coincided with the only gap in LT coverage during the first month, so we lack $H$ and $z$ photometry to confirm its origin.)

\subsection{Host Galaxy Properties}
\label{sec:hostgalaxy}

To characterize the host galaxy in more detail, we gather multi-wavelength photometry from UV to NIR. We use photometry from the NASA Sloan Atlas, which includes both optical photometry from the Sloan Digital Sky Survey \citep[SDSS;][]{York2000} and UV photometry from the Galaxy Evolution Explorer \citep[GALEX;][]{Martin2005} using the {\tt elpetrosian} aperture flux \citep{Blanton2011}. We also perform our own photometry using images from the Pan-STARRS 3pi survey \citep{Kaiser2010}, the Two Micron All-Sky Survey \citep[2MASS;][]{Huchra2012} and the Wide-field Infrared Survey Explorer \citep[WISE;][]{Wright2010}.  Our photometry (AB mags, not corrected for Galactic extinction) is presented in Table \ref{tab:hostphotometry}.

\begin{table}
    \centering
    \caption{Host-galaxy photometry from pre-imaging observations.}
    \label{tab:hostphotometry}
    \begin{tabular}{lccll} % four columns, alignment for each
        \hline
Filter  &   AB mag   &   uncertainty  & Survey \\
\hline
FUV & 18.376 & 0.210 &  GALEX \\
NUV & 17.880 & 0.038 &  GALEX \\
u   & 16.763 & 0.036 &  SDSS  \\
g   & 15.578 & 0.003 &  SDSS  \\
g   & 15.573 & 0.010 &  Pan-STARRS \\
r   & 15.021 & 0.002 &  SDSS  \\
r   & 15.048 & 0.017 &  Pan-STARRS \\
i   & 14.725 & 0.009 &  SDSS  \\
i   & 14.814 & 0.018 &  Pan-STARRS \\
z   & 14.544 & 0.020 &  SDSS  \\
z   & 14.626 & 0.024 &  Pan-STARRS \\
Y   & 14.481 & 0.046 &  Pan-STARRS \\
J   & 14.153 & 0.054 &  2MASS \\
H   & 14.073 & 0.081 &  2MASS \\
Ks  & 14.320 & 0.106 &  2MASS \\
W1  & 15.370 & 0.007 &  WISE  \\
W2  & 16.007 & 0.017 &  WISE  \\
W3  & 14.989 & 0.032 &  WISE  \\
W4  & 14.673 & 0.242 &  WISE  \\
\hline
\end{tabular}
\end{table}

\begin{figure}
\includegraphics[width=\columnwidth]{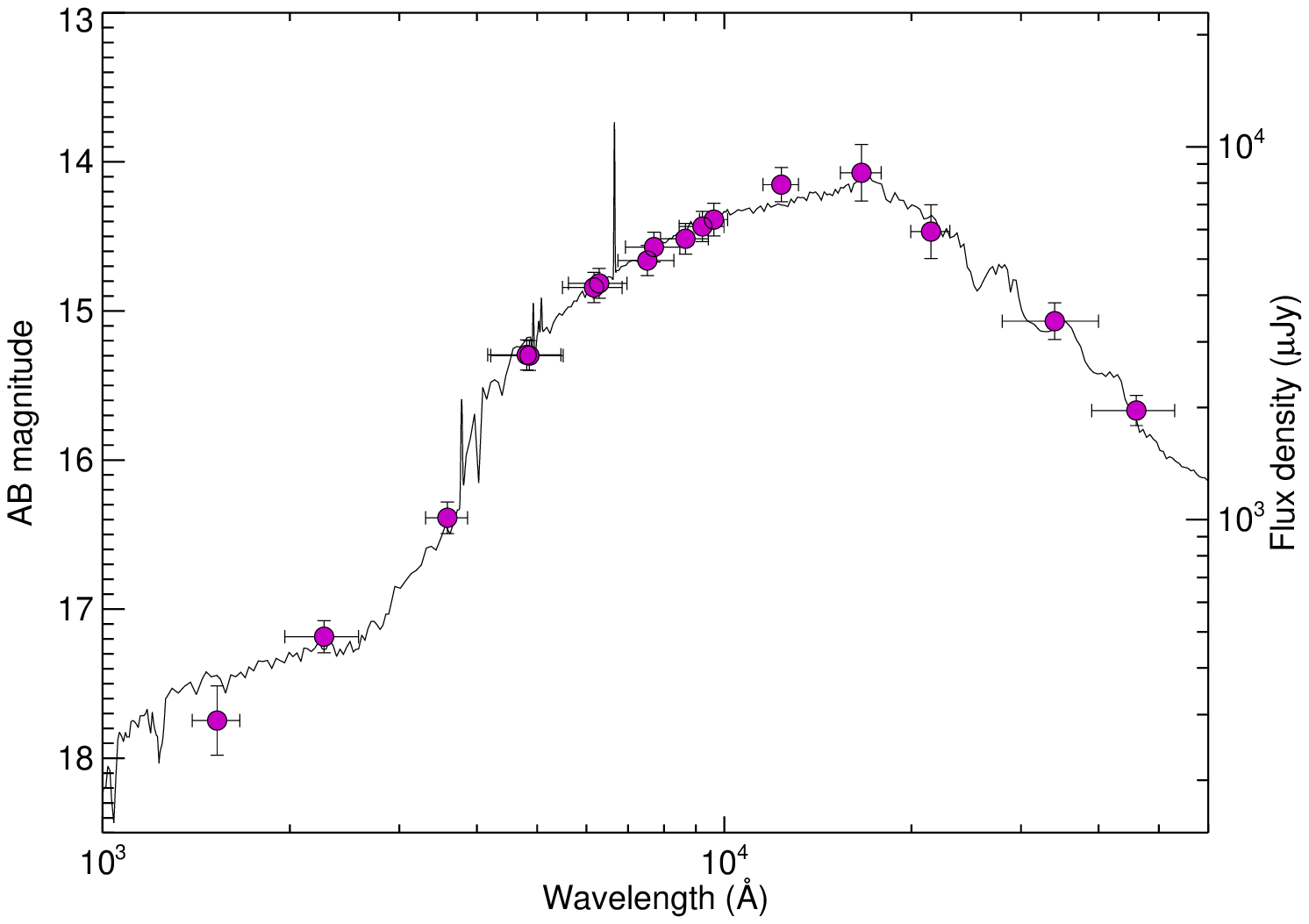}
    \caption{Spectral energy distribution for the host galaxy of AT2018cow. Multi-band photometry from GALEX, SDSS, 2MASS, and WISE is shown in purple and the best-fit SED model (with M$_{*}$ = $1.42^{+0.17}_{-0.29}$ $\times$10$^{9}$ M$_{\odot}$, SFR = 0.22$^{+0.03}_{-0.04}$ M$_{\odot}$ yr$^{-1}$) is shown as a curve.}
    \label{fig:galaxy_sed}
\end{figure}

We fit the broad-band spectral energy distribution using  \verb Le \verb Phare \ \citep{Ilbert2006}, correcting for foreground extinction before fitting the SED.  We assume a \cite{Chabrier2003} IMF, a metallicity between $0.2Z_\odot < Z < Z_\odot$, a \cite{Calzetti2000} extinction law, and otherwise use an identical procedure to that employed in Taggart et al. 2018 (in prep).  The W3 and W4 filters (which are dominated by PAH emission features) were not included in the fit.  We derive a stellar mass of M$_{*}$ = $1.42^{+0.17}_{-0.29}$ $\times$10$^{9}$ M$_{\odot}$ and a total star-formation rate of SFR = 0.22$^{+0.03}_{-0.04}$ M$_{\odot}$ yr$^{-1}$.  The galaxy photometry and final SED fit are shown in Figure~\ref{fig:galaxy_sed}.

These properties suggest a star-forming dwarf spiral similar to the Large Magellanic Cloud.  Its mass is smaller than that of the majority of galaxies that produce core-collapse supernovae, but is well within the distribution. While clearly star-forming, the galaxy is not particularly young nor is it undergoing a notable burst of star-formation.

\section{Comparisons to Previous Events}
\label{sec:comparisons}

\subsection{A Fast Extreme-Luminosity Transient Seen Up-Close}

\begin{figure*}
\includegraphics[width=7in]{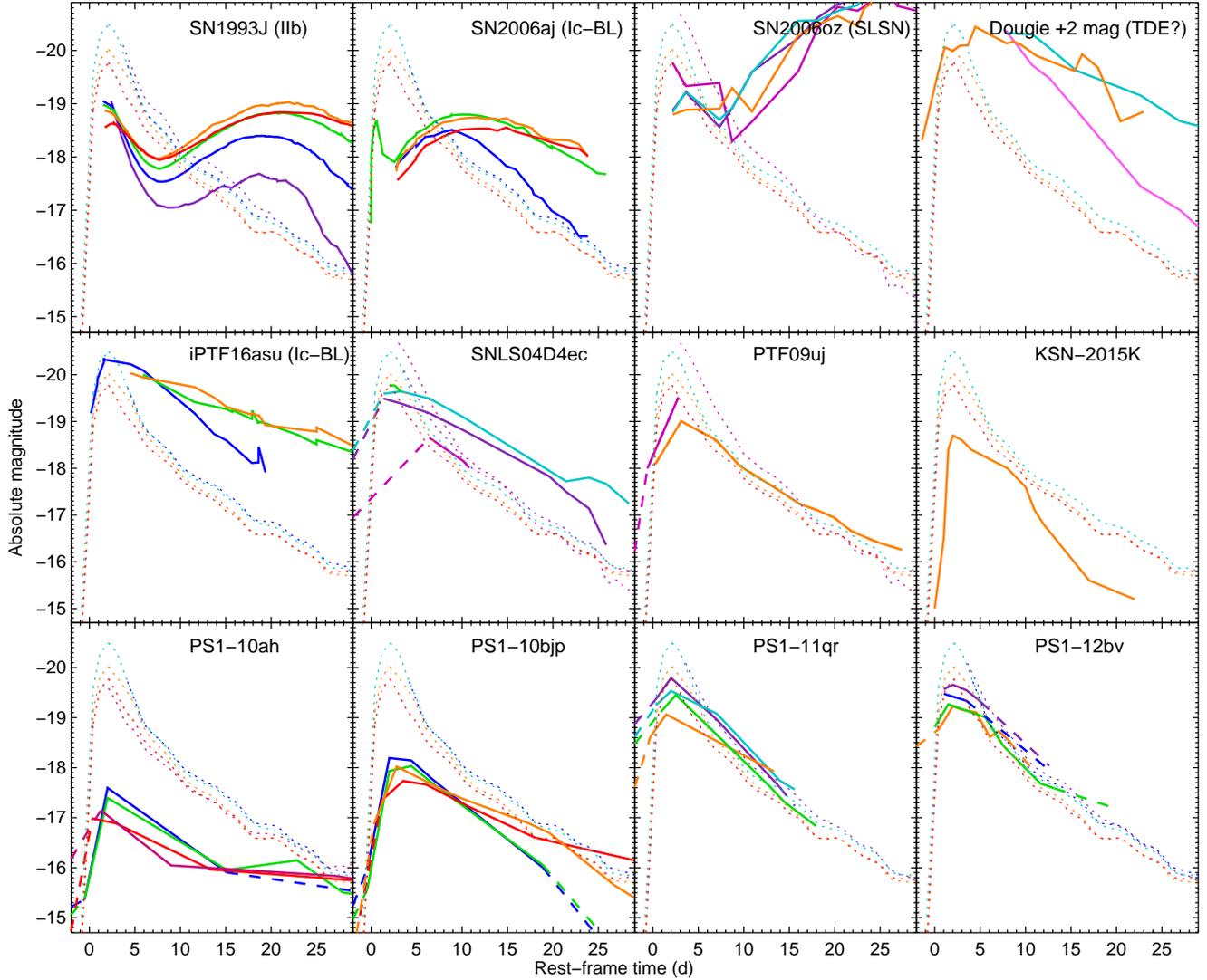}
    \caption{Comparison of the light curve of AT2018cow to other classes of fast-rising or luminous events.  The dotted lines show AT2018cow; solid lines indicate comparison objects with dashed lines connecting upper limits to detections.  Filter colour codes are the same as in Figure \ref{fig:lc} and are matched to rest-frame wavelengths.  AT2018cow is insufficiently luminous at late times compared to GRB-SNe and also far too blue.  It is much faster than any known TDE and the thermal SED is entirely unlike the optically thin spectra of GRB afterglows.  However, it matches well with the cosmological fast transients found in PS1 and Kepler (and to a lesser extent SNLS) in colour, luminosity, and timescale.}
    \label{fig:comparelc}
\end{figure*}

The fast rise, early peak, and subsequent rapid decay do not resemble any common class of extragalactic transient.  While supernovae can show early, luminous peaks associated with shock heating, these are inevitably followed by either a long plateau (as in SNe\,IIP or IIn) or by a second, radioactively powered peak (in SNe IIb, SNe\,Ib/c, and GRB-SNe).  A few classical examples of this are shown in the top row of Figure \ref{fig:comparelc}: SN 1993J \citep{Richmond+1996,Barbon+1995} and SN 2006aj \citep{Campana+2006,Ferrero+2006}, as well as the double-peaked superluminous supernova SN\,2006oz \citep{Leloudas+2012}.  In all cases the late-time flux of these reference objects exceeds that of AT\,2018cow by several magnitudes.

The rest of Figure \ref{fig:comparelc} shows comparisons between the light curve of AT\,2018cow and a variety of luminous, fast-rising transients from different surveys.  These transients are diverse, exhibiting differences in both temporal and colour evolution.  Several retain a high luminosity for a long period and fail to replicate the fast fading of AT\,2018cow.  These include iPTF16asu \citep{Whitesides+2017}, an initially featureless transient that later developed into a SN\,Ic-BL; all members of the \cite{Arcavi+2016} sample (SNLS04D4ec, the fastest of these, is shown); and the unknown transient ``Dougie'' \citep{Vinko+2015}.

The most convincing matches by far are the luminous members of the PS1 sample from \cite{Drout+2014}: PS1-11qr and PS1-12bv, shown at bottom right. While not quite as luminous or as fast-evolving as AT\,2018cow, these events manage to replicate the fast rise, fast decay, and consistent blue colours around the peak time.  (The less luminous objects in that sample are more questionable: in addition to being less luminous by a factor of $\sim$10 they fade more slowly and clearly become redder at late times.)\footnote{The DES fast transients of \cite{Pursiainen+2018} do not yet have publicly available light curves and are not shown in Figure \ref{fig:comparelc}.  Like the PS1 transients, they exhibit a variety of luminosities but all are fast-evolving and most are blue at peak.  Some also show evidence of sustained high temperatures and contracting photospheres, similar to what observed in AT\,2018cow.  The HSC transients of \cite{Tanaka+2016} were observed only in $g$ and $r$ and generally only during the rising phase, so post-peak constraints are not available.}   

Additionally, both PTF\,09uj \citep{Ofek+2010} and KSN-2015K \citep{Rest+2018} also represent good light-curve matches to AT\,2018cow.  Neither has multi-epoch colour information and they are 1--2 mag fainter at peak, although the pre-peak UV-optical colour of PTF\,09uj and the single-epoch colours of KSN-2015K suggest that these transients were indeed similarly blue.

None of these transients have been characterized in detail, although the few spectra that exist are generally featureless (PTF\,09uj exhibited weak, narrow emission lines of hydrogen.)  All were found in star-forming galaxies offset from their host nuclei.  

The rate of fast, blue transients was estimated from the Pan-STARRS sample \citep{Drout+2014}: they measured a value of 4--7\% of the core-collapse supernova rate, equivalent to 1 per year within a radius of 40 Mpc.   Given this rate, it seems credible that one might be detected at 60 Mpc in the first few years of high-cadence all-sky observations by ATLAS or ZTF.  (Conversely, given the detection of an event this close within ATLAS/ZTF, it would be surprising if similar events were \emph{not} present in PS1 and other surveys.)

For these reasons, we argue that AT\,2018cow is very likely related to the population of fast, blue, luminous transients seen by PS1 (and also by DES and HSC; \citealt{Pursiainen+2018,Tanaka+2016}).  Earlier studies almost universally attributed these transients to supernovae undergoing shock breakout into, or interaction with, a dense wind or shell close to the progenitor \citep{Ofek+2010}.  The extensive additional observations available for AT\,2018cow allow us to examine this connection in much more detail.

\subsection{A Spectroscopically Unique Transient}

AT\,2018cow shows at least two distinct spectral phases.  Prior to 10 days it is effectively featureless, save for the short-lived, broad blue absorption feature.  After 12 days it remains hot and blue but exhibits weak features of (redshifted) H, He, and other light elements in emission.

\begin{figure*}
\includegraphics[width=6.5in]{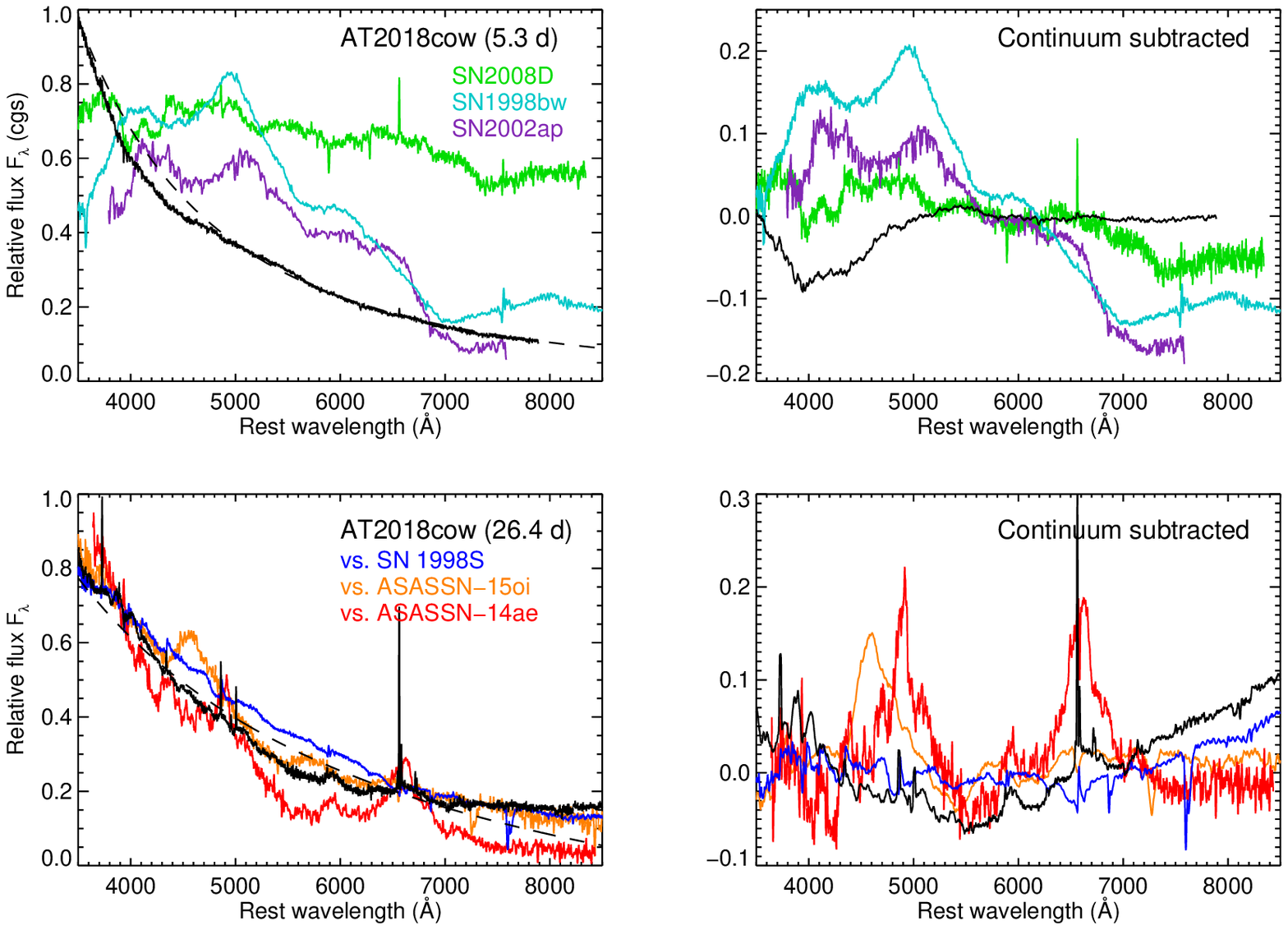}
    \caption{Comparison of an early spectrum of AT\,2018cow to the early spectrum of SN\,2008D (Ib) and two SNe Ic-BL (top row), and of a later-time spectrum to SN\,1998S (IIn) and two tidal disruption events (bottom row).  The extremely blue, smooth continuum bears little resemblance to SN Ib/c, even after attempting to subtract the blue continuum.  (A power-law plus a constant have been subtracted from AT\,2018cow; the reference SN spectra have been subtracted by a constant only.  The strength of the features in the SNe after subtraction has been suppressed by a factor of 2.)  The later-time spectra are dominated by weak emission features of hydrogen and helium; these features are also present in IIn SNe but are much narrower compared to what is seen in AT\,2018cow.  These features are seen in known TDEs with similarly broad widths, although typically much greater strengths.}
    \label{fig:comparespec}
\end{figure*}

The early, broad feature\footnote{We emphasize that the existence of this feature is secure: it is seen with a consistent shape and consistent temporal evolution in at least three different independently-reduced spectrographs (SEDM, SPRAT, HFOSC) and is also evident in our photometry via the evolution of the $B-V$ and $g-r$ colours.} has no obvious analogue in any previous event.  It bears some loose resemblance to the Fe II P-Cygni absorption trough seen in SNe\,Ic-BL, but overlying a much hotter continuum.  We attempted to subtract the hot continuum to test this connection more rigorously, but the match is poor, being both too blue and too broad (Figure \ref{fig:comparespec}) compared to even the earliest spectra of SN1998bw or SN2002ap \citep{Patat+2001,Kinugasa+2002}, or of the spectrum of SN2008D \citep{Modjaz+2009} during its shock-cooling phase.\footnote{This cannot be because the SN features are washed out by a bright afterglow, as was the case for early spectra of SN\,2003dh / GRB 030329 (e.g., \citealt{Hjorth+2003, Deng+2005}): the continuum is far too blue to be predominantly synchrotron in origin (\S \ref{sec:physevol}).}   As of yet we have no convincing explanation for the origin of this feature, other than that it implies very fast (nearly relativistic) ejecta.

The identities of the features seen in later spectra (H and He in broad emission) are secure.  In spite of this, these later spectra bear no obvious resemblance to any class of known supernova.  The strongest similarities are to Type IIn supernovae (which can also remain hot for several weeks after explosion, and are emission-dominated by definition): in the bottom panel of Figure \ref{fig:comparespec} we plot AT\,2018cow versus an early spectrum of SN\,1998S from \cite{Fassia+2001}, which shows a similar blue continuum and most of the same H and He transitions.  However, the lines in AT\,2018cow are not narrow for most of their evolution ($v \sim 6000$ km/s, versus a few hundred km/s for SN\,1998S).  Thomson scattering within ionized matter could broaden a line enough to wash out the narrow component, but this would not produce the net redshift in the emission component that we observe.  The H and He thus must be in the ejecta itself (and seemingly preferentially in receding ejecta given the net redshift).

In fact, the best spectroscopic analogues to AT\,2018cow are not supernovae at all. Our spectra bear a striking resemblance to tidal disruption events: the high temperatures, presence of helium and hydrogen features in emission, and moderate velocities all match what is observed for TDEs.  The spectral features in AT\,2018cow are substantially weaker than in the examples of TDEs that we are aware of (bottom panel of Figure \ref{fig:comparespec}; comparison spectra are from \citealt{Holoien+2014,Holoien+2016}), but the resemblance to a TDE is much stronger than to any supernova.

We summarize the key observational features of AT\,2018cow in Table \ref{tab:properties}.

\section{Interpretation}
\label{sec:interpretation}

\subsection{Supernova Models: A Jet from a Failed Supernova?}

The location of AT2018cow, and its apparent connection to other cosmological events that have also been found outside the nuclei of their host galaxies, give ample justification to consider a supernova as the most natural interpretation of this event.  However, the observational aspects of this event impose severe constraints on any type of stellar explosion.

The first problem for any supernova model is the need to explain the fast rise.  Heating from radioactive \Nifs\ certainly cannot produce it: at least 5 $M_\odot$ of Nickel would be needed to power the luminosity of AT\,2018cow at peak, which is orders of magnitude greater than the total ejecta mass that would be inferred from the fast rise given standard assumptions about diffusion ($M_{\rm ej} \sim {(\frac{t_{\rm rise}}{20 d})}^2 M_\odot$ or approximately 0.01 $M_\odot$; \citealt{Arnett1982,Rest+2018}).

A natural alternative is shock heating. Most core-collapse SNe are believed to exhibit an early shock-breakout and shock-cooling phase in which the stellar photosphere is nearly-instantaneously heated to X-ray temperatures by the emergence of the SN shock, producing a rapid rise in the light curve \citep{Waxman+2017}.  However, for standard types of stellar progenitor the shock-cooling rise time is far \emph{too} fast to explain AT\,2018cow's 2--3 day rise.  A multi-day rise could be achieved only if the progenitor was quite extended ($R \sim 10^{14}$ cm, or about 10 AU).

This radius is similar to that of the largest red supergiants.  However, a massive stellar envelope of this nature would greatly slow down the later evolution of the SN, producing a ``plateau'' phase rather than sudden fading.  The photosphere at the time of shock breakout thus would have to be unbound, with the shock breaking out into a dense wind or ejected shell associated with recent, intense mass loss.

Evidence has been accumulating in recent years that extreme mass loss shortly before explosion is common \citep{Ofek+2014,GalYam+2014,Yaron+2017}, so this may not be surprising.  However, other observations place further strong constraints on the nature of this recent mass loss: the lack of any flash-ionization features, the lack of shocked hydrogen or helium, and the lack of further rebrightenings in the light curve all require that the CSM shell be quite localized in extent.  This may also be possible, if the previous mass-loss episode is both singular and explosive.  

Further constraints on the explosion can be imposed based on the lack of a second, radioactively powered peak in the light curve.  Using the bolometric luminosity at 20 days and scaling relative to SN\,2002ap \citep{Mazzali+2002,Foley+2003}, we estimate $M_{\rm Ni} < 0.05 M_\odot$.  While this is in the range of masses inferred for ``normal'' core-collapse supernovae \citep{Rubin+2016,Muller+2017}, a modest \Nifs\ mass seems hard to reconcile with the energetic shockwave necessary to produce the extraordinary shock-breakout flash and accelerate substantial ejecta to $>0.1$c, as inferred from the broad absorption seen in the spectra at $\sim 1$ week\footnote{It could be contested whether the broad feature truly represents Doppler-broadened absorption, given the lack of a clear identification of the line(s) responsible.  However, as the SEDs in Figure \ref{fig:sedsequence} make clear, this feature shows up clearly as missing flux from what is otherwise an excellent fit to a single thermal SED; multiple emission components or non-thermal features cannot reproduce this profile.   Alternative, non-velocity-broadened sources of absorption (e.g. transient dust extinction with an unknown broad feature) are unlikely.} and by the luminous radio counterpart.  (Velocities this high have been previously seen observationally only in GRB-SNe, which have universally high ejecta and nickel masses: \citealt{Mazzali+2014}, although c.f. \citealt{Fynbo+2006}). 

Perhaps the shock in this SN was driven not by the classical neutrino mechanism (or other forms of energy input from a proto-NS), but solely by an energetic jet driven by a black hole following direct collapse of a massive star to a black hole (analogous to the original ``failed supernova'' model of \citealt{Woosley1993}).  No high-energy prompt emission was observed from AT\,2018cow, but the jet could have been off-axis or (more likely) choked by the stellar envelope.  We may then just have seen a short-lived high-velocity pseudo-photosphere in the early spectra, which may be supported by a small amount of material surrounding the jet, either dragged by the jet itself or ejected in a disc wind. This material would contain only a small amount of \Nifs, explaining the lack of a radioactive second peak.

This model (which is similar to that of \citealt{Quataert+2012}, but with the addition of circumstellar interaction: see also \citealt{Kashayama+2015}) has some appeal, especially given the observation of bright, self-absorbed radio emission which independently implies substantial interaction \citep{Ho+2018}.  Even so, it faces formidable challenges.  The high-velocity absorption implied by our early spectra suggests material that is expanding outward rapidly ($>0.1$c), but the spectral features seen only two weeks later are quite narrow ($\sim0.02$c).  This could be achieved if the high-velocity ejecta collided with a \emph{second} dense shell of comparable mass---eliminating the broad lines and largely halting the expansion of the photosphere that would normally be expected in a young supernova.  But the resulting shock-wave should then have excited narrow-line emission of H and perhaps He which we do not see.  (The H and He lines that eventually emerge originate too late and have velocities too broad to be attributed to shock interaction).  

Alternative stellar progenitor scenarios beyond core-collapse do not provide any appreciable resolution to these contradictions.  Large energies and small \Nifs\ masses are expected for neutron star merger models, but such events should not possess significant hydrogen or helium. Furthermore, AT2018cow empirically bears no relation to the (much dimmer, fast-cooling, fast-expanding) optical counterpart of GW\,170817 \citep{McCully+2017,Kasliwal+2017,Evans+2017,Villar+2017,Pian+2017}.  White-dwarf explosions (variants on Ia or accretion-induced collapse models; e.g., \citealt{Brooks+2017,Poznanski+2010}) are also likely to be poor in H and He, and heavily suppress the UV via iron line blanketing in the ejecta.

Perhaps the biggest challenge for any supernova model is the lack of expansion of the photosphere.  \cite{Pursiainen+2018} noted that a hot, receding photosphere is expected in the wind shock-breakout model due to the rapid expansion of the unbound shock-heated material, but this will only be true during the early phases: the photosphere should eventually reach the dense stellar envelope, after which its evolution should follow that of typical supernovae.   Regardless of the progenitor structure, it is difficult to understand how freely-expanding ejecta would maintain a photosphere on a scale of only $10^{14}$ cm 40 days after the explosion: the material at the photosphere could be expanding no faster than 300 km/s (much slower than the width inferred by the observed lines at late times.)   

\subsection{Tidal Disruption Models: Disruption of a Star by an IMBH?}
\label{sec:tdemodel}

In spite of the circumstantial evidence for a SN origin (the event occurred in a spiral arm) there are many reasons to look more broadly at progenitor models, and in particular to consider a tidal disruption event as an alternative.  

Many of the properties of the transient that cause the most difficulty for the SN interpretation are natural components of TDE models.  The bolometric light curve declines as a power-law, as expected under simple TDE models (although the decay is steeper than the canonical $t^{-5/3}$).  The lack of an early free-expansion phase and the maintenance of a high temperature are also similar to expectations for TDEs, which provide continued energy input via BH accretion and whose potential well hampers free expansion of the ejecta.   And a TDE origin would also explain the H and He-rich late-time spectra (which empirically resemble known TDEs more closely than any SN).

Aside from its peculiar location, the primary feature that distinguishes AT2018cow from known TDEs is its timescale: typical TDEs have rise times of weeks to months and decay times even longer.  Faster TDEs have been found more recently \citep{Blagorodnova+2017}, but even these have characteristic timescales an order of magnitude longer than AT\,2018cow.

A possible resolution is a smaller black hole mass: known TDEs appear to show an empirical timescale-mass correlation (e.g., \citealt{Blagorodnova+2017}), and there are also reasons to expect one theoretically \citep{Guillochon+2013}.  To better constrain the black hole mass under a TDE model, we fit the UV/optical data using two different methods: using simple scaling relations, and using a full MCMC fit to the light curve.

We first fit the bolometric (UVOIR) light curve to a power-law decay of the form $L(t) = L_0(\frac{t-t_0}{t-t_D})^{-n}$.  We obtain an excellent fit with a power-law index of $n = 3.0 \pm 0.1$ and a time of disruption ($t_D$) of $-1.5 \pm 0.3$ (relative to MJD 58285).  Under this scenario the implied rise-time-to peak of $t_{\rm peak} = t_0 - t_D = 5.0 $d, according to the simulations of \cite{Guillochon+2013} for a solar-type star, would correspond to a black hole of $1.5 \times 10^4 M_\odot$.

\begin{figure}
\includegraphics[width=\columnwidth]{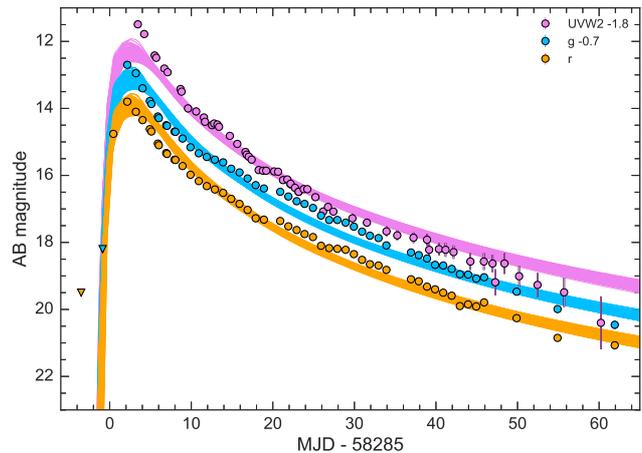}
    \caption{Results from an MCMC fit to the data using the TDE implementation of MOSFiT \citep{Guillochon+2018,Mockler+2018}.  The rising and falling timescales of this transient, along with the slow temperature evolution, are well-reproduced by a model involving the disruption of a Solar-type star around an intermediate-mass black hole ($\sim2\times10^4 M_\odot$).}
    \label{fig:tdemodel}
\end{figure}

Additionally, we fit the light curve in the $g$, $r$, and UVOT $w2$ bands with the MOSFiT TDE model \citep{Guillochon+2018,Mockler+2018}.  The MOSFiT TDE model uses hydrodynamic simulations of tidal disruption events from \citep{Guillochon+2013} to calculate the fallback rate of stellar debris to the black hole. MOSFiT then converts these fallback rates into bolometric luminosities and passes them through viscosity and reprocessing transformation functions to create optical and UV light curves.  Two adjustments to the model were required to obtain a good fit:  the peak luminosity was allowed to exceed the Eddington limit, and the maximum photosphere radius was allowed to reach beyond the apocentres of the Keplerian orbits of the stellar debris.  Under these circumstances, our fit prefers a black hole with a mass of $M_{\rm h} = 1.9^{+1.2}_{-0.8} \times 10^4 M_{\odot}$ and a star with mass $M_{\rm \ast} = 0.6^{+2}_{-0.5} M_{\odot}$.  This is fully consistent with the scaling-relation solution.  Fitted light curves are presented in Figure \ref{fig:tdemodel}.

These parameters correspond to the disruption of a main-sequence star around an intermediate-mass black hole (IMBH).  This would be a significant discovery: IMBH disruptions were recently theoretically predicted \citep{Fragione+2018,Chen+2018} and if confirmed, would represent evidence for the existence of IMBHs in the low-redshift universe, a topic that remains broadly controversial.  A black hole in this mass range would also not conflict with the off-nuclear location: it could originate from a globular cluster or from a massive young star cluster.

However, as the above discussion suggests, the peak luminosity of the transient ($\sim3\times10^{44}$ erg) is much greater than the Eddington luminosity for a black hole of the mass needed to explain its short timescale ($\sim10^{42}$ erg for $M_{\rm BH} = 10^4 M_\odot$).  While TDEs are expected to have super-Eddington mass fallback rates (e.g., \citealt{Strubbe+2009}), the radiated luminosity is generally expected to be capped at close to the Eddington luminosity \citep{Chen+2018}, since higher luminosities would disrupt the accretion and drive the luminosity back down. Super-Eddington luminosities could be achieved in two ways: by an anisotropic radiation process, or by a heating source not directly associated with accretion.  

There is evidence that some TDEs can indeed produce highly anisotropic, relativistic jets \citep{Bloom+2011,Levan+2011,Cenko+2012b,Burrows+2011}.  The bright (and variable) X-ray and radio emission from AT\,2018cow (see also \citealt{Ho+2018,Margutti+2018}) suggest a similar phenomenon could be present here as well.  However, the optical radiation which gives rise to our luminosity estimates is unambiguously thermal and not easily beamed, so anisotropy is unlikely to resolve the conflict. 

Alternatively, it is possible that the early UV/optical emission is related to the circularization process \citep{Piran+2015,Dai+2018}, rather than accretion.  
The similarity of the peak luminosity of AT2018cow to other UV/optical TDEs \citep{Hung+2017} and the expected energy dissipation rate from the circularization process of $10^{44} (M_{\rm BH}/10^6 M_{\odot})^{-1/6}$ erg s$^{-1}$ \citep{Piran+2015} support this interpretation.
The self-intersection radius for debris streams around a $~10^4$ $M_\odot$ black hole is $\sim 5 \times 10^{13}$ cm \citep{Wevers+2017}, which is a factor of 10 smaller than the observed photosphere radius for AT2018cow. If the luminosity is powered by stream-stream intersections, then the photosphere would engulf both the intersection point and the black hole. This optically thick reprocessing layer would need to be in place by the time of our first observations to explain the colour and luminosity of AT2018cow.  This could be associated with matter blown to larger radii during an early wind phase \citep{Jiang+2016,Metzger+2016}.

Further modeling will be needed to examine the behaviour of tidal disruptions around IMBHs during the super-Eddington phase.  If even some of the PS1 and DES events belong to the same class as AT\,2018cow, there is reason to believe that these events are reasonably common and the current generation of fast-cadence optical surveys may find future examples at similar rates as ordinary, SMBH TDEs.\footnote{Super-Eddington-luminosity disruptions by more massive black holes are also of interest: the transient ``Dougie'' was slower than AT\,2018cow (Figure \ref{fig:comparelc}) but was vastly more luminous, and the preferred TDE model fit by \cite{Vinko+2015} also indicated a highly super-Eddington luminosity, in this case from a somewhat more massive black hole ($M_{\rm h} = 2.0^{+13.9}_{-1.3} \times 10^5 M_{\odot}$).}

\section{Conclusions}
\label{sec:conclusions}

Prior to AT\,2018cow, fast high-luminosity transients were widely attributed to an extreme variant of the shock-breakout scenario that has already been widely appealed to in order to explain a variety of nearby supernovae.
To our surprise, the first real-time detection of a nearby event belonging to this empirical class has only deepened the mystery surrounding these events.  While the off-nuclear location within a star-forming region seems to imply the explosion of a star as a supernova, the actual observational properties---including high-velocity absorption in early spectra, a long-lived hot photosphere, a complete lack of narrow lines during the first week, and luminous X-ray through radio emission--- are all difficult to explain under any existing supernova model.  If nothing else, any stellar explosion must involve a radically different progenitor structure and/or explosion mechanism compared to known SNe.  

In contrast, disruption by an intermediate-mass black hole provides an excellent description of the qualitative behaviour of the transient and its later-time spectra.   However, the highly super-Eddington luminosity of the transient is a formidable challenge for IMBH TDE models, and it remains to be seen whether alternative explanations for the early heating (e.g. circularization of infalling material) provide an adequate explanation.

Studies of fast optical transients are still in their infancy, and there is much more to learn both observationally and theoretically.  While an event as close as AT\,2018cow may not be a regular occurrence, its sheer brightness suggests that others of a similar nature are likely to be observed in the near future at somewhat greater distances.  Samples of the spatially-resolved galaxy environments, total energetics, and spectroscopic properties of such events are likely to shed light on their nature.

\section{Acknowledgments}
DAP acknowledges useful discussions with Matt Darnley and Helen Jermak.  We thank the referee for helpful comments which improved the quality of this paper.

This work was supported by the GROWTH project funded by the National Science Foundation under Grant No 1545949. GROWTH is a collaborative project between California Institute of Technology, Pomona College, San Diego State University, Los Alamos National Laboratory, University of Maryland College Park, University of Wisconsin Milwaukee (USA), Tokyo Institute of Technology (Japan), National Central University (Taiwan), Indian Institute of Astrophysics (India), Weizmann Institute of Science (Israel), The Oskar Klein Centre at Stockholm University (Sweden), Humboldt University (Germany), and Liverpool John Moores University (UK).  This paper used the GROWTH marshal to filter alerts and co-ordinate follow-up.

FT and JS gratefully acknowledge support from the Knut and Alice Wallenberg Foundation. JS acknowledges the support of Vetenskapsr\aa det through VR grants 2012-2265 and 2017-03699. 

AYQH was supported by a National Science Foundation Graduate Research Fellowship under Grant No. DGE-1144469.
MC was supported by the David and Ellen Lee Postdoctoral Fellowship at the California Institute of Technology. 
JSB was supported by a Data-Driven Discovery grant from the Moore Foundation.

C.-C.~Ngeow and P.-C.~Yu thank the Ministry of Science and Technology (MoST, Taiwan) for funding under grant 104-2923-M-008-004-MY5 and 106-2112-M-008-007. This publication has made use of data collected at Lulin Observatory, partly supported by MoST grant 105-2112-M-008-024-MY3.

RI is supported by JSPS and NSF under the JSPS-NSF Partnerships for
International Research and Education (PIRE),
YT and NK supported by JSPS KAKENHI Grant Numbers JP16J05742 and JP17H06362.
YT is also financially supported by Academy for Global Leadership
(AGL) of Tokyo Institute of Technology.
MITSuME Akeno 50cm telescope is also supported by the joint research
program of the Institute for Cosmic Ray Research (ICRR),
the University of Tokyo, and Optical and Near-Infrared Astronomy
Inter-University Cooperation Program in Japan (KLM).

GCA and VB acknowledge the support of the Science and Engineering Research Board, Department of Science and Technology, India and the Indo-US Science and Technology Forum for the GROWTH-India project. BK acknowledges the Science and Engineering Research Board under the Department of Science \& Technology, Govt. of India, for financial assistance in the form of National Post-Doctoral Fellowship (PDF/2016/001563) and BRICS grant
DST/IMRCD/BRICS/PilotCall1/MuMeSTU/2017(G).

The Liverpool Telescope is operated on the island of La Palma by Liverpool John Moores University in the Spanish Observatorio del Roque de los Muchachos of the Instituto de Astrofisica de Canarias with financial support from the UK Science and Technology Facilities Council.  We acknowledge helpful support from the entire LT staff, including Robert Smith, Jon Marchant, and Iain Steele, and to the LT review panel for approving our requests for Reactive time (JQ18A01).  

We acknowledge the use of public data from the Swift data archive, and thank the Swift team for executing such a thorough observing campaign of this transient.

This work is partly based on observations made with the Nordic Optical Telescope, operated by the Nordic Optical Telescope Scientific Association at the Observatorio del Roque de los Muchachos, La Palma, Spain, of the Instituto de Astrofisica de
Canarias. The data presented here were obtained in part with ALFOSC, which
is provided by the Instituto de Astrofisica de Andalucia (IAA) under a joint
agreement with the University of Copenhagen and NOTSA.  The William Herschel Telescope is operated on the island of La Palma by the Isaac Newton Group of Telescopes in the Spanish Observatorio del Roque de los Muchachos of the Instituto de Astrof\'isica de Canarias. 

These results made use of the Discovery Channel Telescope at Lowell Observatory. Lowell is a private, non-profit institution dedicated to astrophysical research and public appreciation of astronomy and operates the DCT in partnership with Boston University, the University of Maryland, the University of Toledo, Northern Arizona University and Yale University. The upgrade of the DeVeny optical spectrograph has been funded by a generous grant from John and Ginger Giovale. We thank Debra Fischer for graciously assisting with our Target of Opportunity request at the DCT.

This work is partly based on observations made with the Kitt Peak EMCCD Demonstrator (KPED) camera on the Kitt Peak 84 inch telescope. The KPED team thanks the National Science Foundation, discretionary funds of SRK, and donors to SRK for support in the building and operation of KPED. In addition, they thank the Chimera project for use of the EMCCD.

Some of the data used in this paper were acquired with the COATLI telescope and interim instrument, and with the RATIR instrument on the 1.5-meter Harold L. Johnson telescope; both at the Observatorio Astron\'omico Nacional on the Sierra de San Pedro M\'artir, Baja California, M\'exico. COATLI is funded by CONACyT (LN 232649, 260369, and 271117) and the Universidad Nacional Aut\'onoma de M\'exico (CIC and DGAPA/PAPIIT IT102715, IG100414, and IN109408).  RATIR is funded by the University of California and NASA Goddard Space Flight Center.  COATLI and the Johnson 1.5-m telescope are operated and maintained by the Observatorio Astron\'omico Nacional and the Instituto de Astronom{\'\i}a of the Universidad Nacional Aut\'onoma de M\'exico. We acknowledge the contribution of Leonid Georgiev and Neil Gehrels to the development of RATIR.

Based partially on data from the Gemini Observatory, which is operated by the Association of Universities for Research in Astronomy, Inc., under a cooperative agreement with the NSF on behalf of the Gemini partnership: the National Science Foundation (United States), the National Research Council (Canada), CONICYT (Chile), Ministerio de Ciencia, Tecnolog\'{i}a e Innovaci\'{o}n Productiva (Argentina), and Minist\'{e}rio da Ci\^{e}ncia, Tecnologia e Inova\c{c}\~{a}o (Brazil).  Data were taken under program GN-2018A-Q-902 and acquired through the Gemini Observatory Archive.  

Some data presented herein were obtained at the W. M. Keck Observatory, which is operated as a scientific partnership among the California Institute of Technology, the University of California and the National Aeronautics and Space Administration. The Observatory was made possible by the generous financial support of the W. M. Keck Foundation. The authors wish to recognize and acknowledge the very significant cultural role and reverence that the summit of Maunakea has always had within the indigenous Hawaiian community.  We are most fortunate to have the opportunity to conduct observations from this mountain. 

The 2m Himalayan Chandra Telescope at the Indian Astronomical Observatory is operated by the Indian Institute of Astrophysics. Support from IAO staff and HCT observation assistants is acknowledged. We also thank all HCT observers who provided part of their time for the observations.  The HCT data of 22, 23, 25 and 29 June were obtained under the ToO proposals of both D.K.Sahu (PI) and F. Sutaria (PI).

AstroSat is a dedicated multi-wavelength space observatory funded and facilitated by the Indian Space Research Organisation (ISRO). We thank the AstroSat ToO Time Allocation Committee for granting ToO time, and the operations team for carrying out these observations. 

Funding for the SDSS and SDSS-II has been provided by the Alfred P. Sloan Foundation, the Participating Institutions, the National Science Foundation, the U.S. Department of Energy, the National Aeronautics and Space Administration, the Japanese Monbukagakusho, the Max Planck Society, and the Higher Education Funding Council for England. The SDSS Web Site is http://www.sdss.org/.  The SDSS is managed by the Astrophysical Research Consortium for the Participating Institutions.

The Pan-STARRS1 Surveys (PS1) and the PS1 public science archive have been made possible through contributions by the Institute for Astronomy, the University of Hawaii, the Pan-STARRS Project Office, the Max-Planck Society and its participating institutes, the Max Planck Institute for Astronomy, Heidelberg and the Max Planck Institute for Extraterrestrial Physics, Garching, The Johns Hopkins University, Durham University, the University of Edinburgh, the Queen's University Belfast, the Harvard-Smithsonian Center for Astrophysics, the Las Cumbres Observatory Global Telescope Network Incorporated, the National Central University of Taiwan, the Space Telescope Science Institute, the National Aeronautics and Space Administration under Grant No. NNX08AR22G issued through the Planetary Science Division of the NASA Science Mission Directorate, the National Science Foundation Grant No. AST-1238877, the University of Maryland, Eotvos Lorand University (ELTE), the Los Alamos National Laboratory, and the Gordon and Betty Moore Foundation.

Finally, we thank the developers and maintainers of the Open Supernova Catalog and the Weizmann Spectral Repository, whose databases greatly facilitated the comparisons of this event to other transients.

\bibliographystyle{mnras}
\bibliography{ref}

\end{document}